\documentclass[%
 reprint,
superscriptaddress,
 amsmath,amssymb,
pra,
]{revtex4-2}

\usepackage{graphicx}% Include figure files
\usepackage{dcolumn}% Align table columns on decimal point
\usepackage{bm}% bold math
\usepackage{color,soul} % highlight
\usepackage{natbib}

\usepackage[pdftex,plainpages=false,
    pdfpagelabels,
    pdfusetitle,
    bookmarksnumbered,
    linkbordercolor={1 1 1},
    hidelinks=true,
    colorlinks={ true },
]{hyperref}

\usepackage[version=4]{mhchem}

\newcommand{\mz}{\ensuremath{m/z}}

\newcommand{\ie}{i.\,e.}
\newcommand{\eg}{e.\,g.}

\newcommand{\densU}{\ensuremath{\text{cm}^{-3}}}
\newcommand{\VRe}{\ensuremath{V_{\mathrm{R4}}}}
\newcommand{\VeffRe}{\ensuremath{V^{*}_{\mathrm{R4}}}}
\newcommand{\VSE}{\ensuremath{V_{\mathrm{SE}}}}
\newcommand{\VSA}{\ensuremath{V_{\mathrm{SA}}}}
\newcommand{\Vhalf}{\ensuremath{V_{0.5}}}
\newcommand{\Vtenth}{\ensuremath{V_{0.1}}}
\newcommand{\Ekz}{\ensuremath{E_{\mathrm{k, z}}}}
\newcommand{\Etotal}{\ensuremath{E_{\mathrm{total}}}}
\newcommand{\Ek}{\ensuremath{E_{\mathrm{k}}}}
\newcommand{\Hep}{\ce{He+}}
\newcommand{\HHHp}{\ce{H3+}}

\begin{document}

\preprint{APS/123-QED}

\title{Ion-kinetic-energy sampling in a 22-pole trap using ring-electrode
evaporation}% Force line breaks with \\

\author{Miguel Jim{\'e}nez-Redondo}
\email{mjimenez@mpe.mpg.de}
\affiliation{Max Planck Institute for Extraterrestrial Physics,
Gie{\ss}enbachstra{\ss}e 1, 85748 Garching, Germany}
\author{Dieter Gerlich}
\affiliation{Max Planck Institute for Extraterrestrial Physics,
Gie{\ss}enbachstra{\ss}e 1, 85748 Garching, Germany}
\affiliation{Department of Physics, Technische Universit\"{a}t Chemnitz,
09107 Chemnitz, Germany}
\author{Paola Caselli}
\affiliation{Max Planck Institute for Extraterrestrial Physics,
Gie{\ss}enbachstra{\ss}e 1, 85748 Garching, Germany}
\author{Pavol Jusko}
\affiliation{Max Planck Institute for Extraterrestrial Physics,
Gie{\ss}enbachstra{\ss}e 1, 85748 Garching, Germany}

\date{\today}% It is always \today, today,

\begin{abstract}
We present an experimental method for the characterization of the kinetic
    energies of ions confined in a 22-pole radio frequency
    trap by inducing a small potential
    barrier using the surrounding ring electrodes, allowing the selective
    extraction of ions.
    Energy sampling experiments have been performed on buffer gas thermalized
    He$^+$ ions at trap temperatures between $10-180$~K, resulting
    in distinct extraction curves as a function of the potential barrier, and a
    differentiated behavior depending on the escape time from the trap.
    The experiments are complemented by Monte Carlo simulations of the
    ion trajectories inside the calculated trap potential and
    allow us to investigate the properties of the sampling method,
    the role of ion motion coupling,
    and the impact of residual buffer gas collisions
    on the observed results. The technique has also been
    successfully applied to identify energetic \HHHp\ ions produced in
    an exothermic reaction inside the trap.
    Upon calibration, this method can provide
    relative kinetic energy distributions or
    be used to filter the maximum desired
    kinetic energy of the ions inside the trap.
\end{abstract}

\maketitle

\section{Introduction}

Radio frequency (rf) ion traps are useful tools for the study of ion-molecule
processes. The combination of high-order multipole ion traps, providing large
field-free trapping regions, with cryogenic cooling allows the study of such
processes at temperatures relevant for interstellar chemistry. A careful
determination of the temperature, or more generally, the energy distribution of
the trapped ions, is
crucial in order to accurately characterize the
studied process. Furthermore, insight on the energies of the trapped ions can
potentially help with identification of the trapped species, allowing for the
differentiation of isomers or internally excited species.

The measurement of the temperature of the ion ensemble is commonly sufficient
to characterize the ion energies. Thermometry techniques for ion traps include
the use of rotational spectroscopy to determine the rotational temperature and
using the Doppler broadening of the lines to obtain the translational
temperature, using photodetachment
\cite{notzoldSpectroscopyIonThermometry2022,Plasil2023,
endresIncompleteRotationalCooling2017},
laser induced reaction \cite{Jusko2014} or rare gas tagging \cite{Brunken2017}
action schemes. Time-of-flight distributions have also been proposed as a
method for temperature determination \cite{notzoldThermometryMultipoleIon2020}.
In magneto-optical traps (MOT), the temperature of the atoms has been derived
from the free expansion of the ion cloud \cite{Williams2017}, and a more
precise determination of the energy distribution has been obtained by coupling
a single atom standing-wave dipole trap to the MOT \cite{Alt2003}.

Energetic characterization of photodissociation processes has been tackled with
velocity map imaging setups. These have been successfully coupled to
cylindrical \cite{huaCryogenicCylindricalIon2019} and Paul
\cite{johnstonVelocityMapImaging2018} traps in order to determine the kinetic
energy of the photodissociation fragments.

Some works have dealt with the use of electrostatic potentials for energy
characterization in traps. \citet{champeauKineticEnergyMeasurements1994} used a
retarding potential to discriminate ions according to their energy after their
extraction from a quadrupole ion trap and contrasted the results with their
time-of-flight measurements.
Evaporative techniques, based on the use of a potential barrier
to selectively extract ions according to their energy, have been used to
characterize the ions in 22-pole trap experiments.
These methods can be used to sample the energy distribution
of the ions along the different trap axes, and unlike time-of-flight
measurements, can selectively preserve the low energy ions in the trap for
further experiments.
\citet{lakhmanskayaPropertiesMultipoleIon2014}
studied the evaporative ion losses in a 22-pole trap in the radial and axial
directions by regulating the trap temperature and end cap voltage,
and used the loss rate to derive the ion temperature.
This sort of technique is also used
in the recently developed
leak-out spectroscopy (LOS) method
\cite{schmidLeakOutSpectroscopyUniversal2022}, which uses vibrational to
translational energy transfer to generate energetic ions capable of escaping
through the end cap potential.
Forced evaporative cooling has also been successfully applied
to negatively charged particles both in the context of antimatter research
\cite{Andresen2010}, by cooling antiprotons down to 9~K;
and for molecular anions in an octopole trap \cite{Tauch2023}, where
it is demonstrated to be more effective than buffer-gas cooling.

The question of the ion energy distribution in multipole traps has also been
approached from a computational point of view, usually with an emphasis on
buffer gas cooling and the effect of the micromotion induced by the rf field in
the final energy distribution \cite{devoePowerLawDistributionsTrapped2009,
asvanyNumericalSimulationsKinetic2009,
rajeevanNumericalSimulationsStorage2021,
holtkemeierBufferGasCoolingSingle2016,
holtkemeierDynamicsSingleTrapped2016}.
Particularly, the effect of end cap voltage and buffer gas mass has been
investigated using Monte Carlo simulations of the ion trajectories in 22 pole
\cite{asvanyNumericalSimulationsKinetic2009} and 16 pole and 16 wire
\cite{rajeevanNumericalSimulationsStorage2021} traps. Trajectory simulations
have also been used to model the trapping process and the resulting energies in
a ring electrode trap without the use of buffer gas
\cite{svendsenTrappingIonsFast2013}. In the \textmu K regime, the limits of
buffer gas cooling of atomic ions in Coulomb crystals have been explored for
Paul traps \cite{furstProspectsReachingQuantum2018} and then extended to
several trap geometries including multipoles \cite{trimbyBufferGasCooling2022}.

In this work, we present an experimental method for the characterization of the
kinetic energy distribution of ions inside a 22-pole trap.
The small the
potential barrier generated by a surrounding ring electrode,
which allows for finer control over the energy of the ions
compared to the end cap voltage,
is used to discriminate the
ions according to their energy.

This setup is used to characterize
thermalized \Hep\ ions at different trap temperatures.
Monte Carlo simulations of the ion trajectories under energy sampling
conditions are used to reproduce the experimental data and %help with the
support the interpretation of the experimental results.
The method is also applied to \HHHp\ ions produced
inside the trap in an exothermic
reaction, highlighting the potential to selectively sample energetic ions.

\section{Experimental setup}

The experimental setup has been previously described in detail in
\cite{juskoColdCASIon2023}.  It consists of a 22-pole trap (5~cm long, 1~cm
diameter) with stainless steel rods (1~mm diameter) surrounded by five ring
electrodes. The trap is mounted on top of a helium cryostat
which allows the cooling of the device down to 4 K. A
heater cartridge
is used to set the desired temperature
which is measured
using silicon diodes placed on the trap enclosure. The trap temperature
can be determined with an accuracy of $\pm 1$~K in this way. The
pressure in the trap vacuum chamber is monitored using a Bayard-Alpert gauge
calibrated by a capacitance manometer directly connected to the trap volume.
This setup can measure pressures down to $\sim 10^{-8}$~mbar inside the trap,
but cannot be reliably used to monitor changes over short timescales, such as
the He buffer gas pulse used in the experiments.

\Hep\ ions are produced in a Gerlich type Storage Ion Source (SIS)
\cite{gerlichInhomogeneousRFFields1992} by electron bombardment of neutral He,
and are subsequently guided through a quadrupole acting as a mass filter and an
electrostatic bender before arriving into the trap. A short (24~ms)
and intense He pulse, injected through a custom piezo-element actuated valve,
is used to slow down the ion beam which allows efficient trapping
of the ions. This He also acts as a buffer cooling gas, allowing the ions to
cool down to a temperature close to the one of the trap.
After the desired trapping time, ions are released to a second
quadrupole allowing the \mz\ characterization by mass spectrometry
and subsequently detected using a conversion dynode together
with a channel electron multiplier. A multi-channel scaler (MCS) enables
resolving the time of arrival of the ions at the detector.

\begin{figure}
\centering
\includegraphics[width=\linewidth]{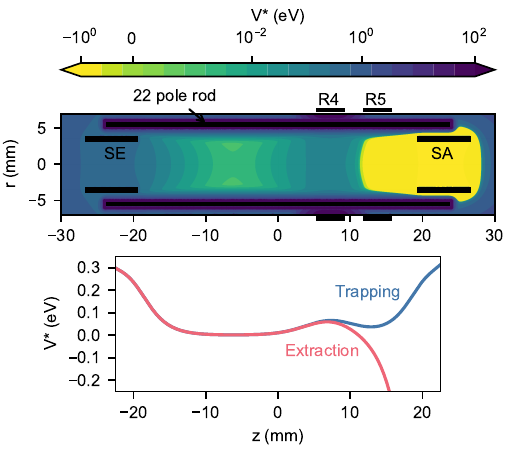}
    \caption{\label{fig:Veff} (Color online)
    Top panel: cut through the calculated effective
    mechanical potential inside the 22-pole trap (\textit{rz})
    set for the extraction
    configuration of the energy sampling experiment.
    The frequency and amplitude of the rf voltage applied to the 22 poles
    are $f_0 = 19\,\text{MHz}$ and $V_0 = 57$~V.
    The voltages for the entrance, exit and ring electrodes are
    $\VSE = 0.3$~V, $\VSA = -1.5$~V and
    $\VRe = 80$~V respectively.
    The corresponding ring electrode barrier height, \VeffRe, is 58.6~meV.
    Bottom panel: effective mechanical potential calculated
    on the trap axis for the configurations employed in the
    experiments when trapping (blue) and extracting (red) the ions. Electrode
    voltages are the same as in the top panel, $\VSA = 0.3$~V for the
    trapping configuration.}
\end{figure}

During each 2~s energy sampling experiment, the trap alternates
between two different configurations in which only the voltage
applied to the exit electrode, SA, changes.
For the full duration of the cycle,
the ring electrode R4 (the second closest to the
trap exit) is set to a voltage \VRe\ between 0 and +147~V,
creating a small
($\sim 1:1000$ ratio) barrier inside the trap.
During loading of the trap
(30~ms) and the subsequent 1~s trapping time,
the exit electrode SA is kept at a
positive value (0.3~V) in order to ensure the storage of the ions.
This is referred to hereafter as the trapping configuration.
The 1~s trapping time ensures ions have enough time to cool down
and allows the neutral He buffer gas
(pulsed into the trap in the few first ms of
the sequence) to be evacuated. After this storage time, the
SA electrode is set for extraction, \ie, to negative voltage ($-1.5$~V),
such that it allows the ions with sufficient energy
to overcome the R4 barrier and be guided out of the trap
towards the detector. The escaped ions are counted in a time-resolved
manner (MCS) in a 300~ms
interval following the opening of the SA electrode.
Each experiment is repeated 60 times in order to obtain a good signal over
the whole detection interval. The calculated effective mechanical potential
$V^*$ inside the trap for both configurations is depicted in Fig.~\ref{fig:Veff}.

The extraction potential created by SA can distort the potential barrier
created by the ring electrodes, making the ring electrode closest to
the trap exit, R5, a poor choice for energy sampling experiments. Therefore,
as we have previously described, we chose to use
the next ring electrode, R4, instead \cite{juskoColdCASIon2023}.
Moreover, the influence of SA is still relevant for the lower R4 voltages
employed, requiring careful examination of this effect for the energy
characterization of the ions.
The simulations also showed that the barrier has a radial profile with the
minimum value at the axis of the trap, which possibly affects the sampling
of the ions \cite{juskoColdCASIon2023}.
In any case, when discussing the potential barrier, we always refer to its
value at the axis, noted by \VeffRe, throughout this work.

\section{Simulations}

In order to interpret the experimental results, we performed Monte Carlo
simulations of a number of ions (usually 1000) moving inside the given trap
potential and potentially colliding with buffer gas atoms. The trajectory of
each ion is solved until either the ion reaches the exit electrode SA or a
total of 100~ms are elapsed.
These simulations were performed for varying values of the ion temperature
(or buffer gas temperature, see below), buffer gas pressure and R4 voltage.

It is critical in these simulations to get an accurate reproduction of the
individual ion energetics as it moves inside the trap. Most standard
integration methods struggle with energy conservation, as the total ion energy
slowly drifts, usually to higher values. Therefore, we have used the velocity
Verlet algorithm \cite{swopeComputerSimulationMethod1982}, belonging to the
family of symplectic methods which ensure low energy drift over extended
simulation times \cite{ruthCanonicalIntegrationTechnique1983}, to solve the ion
trajectories.
A second important factor is the treatment of the field governing
the ion motion.
We adopt the effective mechanical potential $V^*$,
from the Boundary Element Method (BEM)
calculations \cite{smigajSolvingBoundaryIntegral2015,
betckeBemppclFastPython2021}
using a 3D mesh of the trap generated with the Salome platform
\cite{SalomePlatform}
(cf. \cite{juskoColdCASIon2023} for further details),
to solve the ion trajectories, given
that the macromotion of the ions is controlled by $V^*$.
This is useful for
easier visualization of the general properties of ion motion inside the trap.
For most of the simulations, however, a time-dependent rf field has been
employed. In this case, the electric field for an ideal multipole
\cite{gerlichInhomogeneousRFFields1992}
\begin{equation}
\begin{pmatrix}
E_x\\E_y
\end{pmatrix}
    = \frac{V_0}{r_0} n \left( \frac{r}{r_0} \right)^{n-1}
    \begin{pmatrix}
-\cos\left(n-1\right)\varphi\\\sin\left(n-1\right)\varphi
\end{pmatrix}
\cos \Omega t
\end{equation}
where $r$ and $\varphi$ are the polar coordinates, $r_0$ is the inscribed
radius, $n=11$ is the multipole order, and $V_0$ and
$\Omega=2 \pi f_0$ are the
amplitude and angular frequency of the
time-dependent rf voltage applied to the 22 poles respectively,
has been superimposed to the electrostatic contribution of the end
and ring electrodes from the BEM simulations. Attempting to use the BEM
calculated rf field instead resulted in unstable trajectories likely due to
numerical artifacts in low potential regions.

The individual ion trajectories are solved using a fixed time step (required by
the symplectic algorithm) that is chosen at the start of the simulation based
on the total ion energy and recalculated if the energy changes due to a
collision. This step is chosen as large as possible for computational
efficiency while still ensuring that energy is conserved through the
trajectory, and is of the order of ns (in comparison the rf period is 52~ns).
At fixed times during the simulation, namely every \textmu s, a check
is performed on whether the ion has escaped the trap, the trajectory data is
saved, conservation of energy is tested and the probability of an elastic
collision of the ion with the buffer gas is checked using the Langevin cross
section. In case a collision occurs, the outcome is determined following
\citet{londryComputerSimulationSingleion1993}, generating a random scattered
direction for the ion and a random Maxwellian velocity for the neutral and
calculating the resulting ion velocity. Ions are considered to have
escaped from the trap when they reach the axial coordinate of the exit
electrode SA ($z = 20$~mm).
The exit time recorded is the time needed to reach this position,
since the focusing and detection of the ions is not simulated.
The flight time between leaving the trap (SA) and detection event
in the experiment is negligible on these time scales.
Nevertheless, the configuration of the ring electrode could
potentially affect the
experimental focusing of the ions and change the detection efficiency.
A careful treatment of this effect would require extending the simulation
of the ion trajectories up to the detector and is out of the scope of
this work, but needs to be considered when comparing experimental and
simulation results.

The velocity distribution of ions collisionally cooled by a buffer gas in a
22-pole trap diverges from a Maxwell-Boltzmann distribution due to the
potential of the trap, which leads to the depletion of the low energy region of
the distribution, and the `rf heating' effect caused by collisions with the
buffer gas particles in regions with a strong rf field
\cite{asvanyNumericalSimulationsKinetic2009}. A Tsallis distribution
{\cite{tsallisPossibleGeneralizationBoltzmannGibbs1988}} is commonly used to
empirically fit the resulting energy distribution
\cite{notzoldThermometryMultipoleIon2020,
devoePowerLawDistributionsTrapped2009}.
Thus, in order to obtain a good initial energy distribution, the simulation of
the experiment is divided in two steps. In the first one, ions are cooled down
from room temperature by collisions with $10^{-4}$~mbar of buffer gas during
100~ms, which is sufficient to ensure thermalization, keeping the exit
electrode closed. After the simulation is done, the resulting final velocities
and positions of the ions are used as initial conditions for the extraction
simulation, where the voltage applied to the SA electrode is switched to a
negative value.
For simplicity, we will simply refer to these conditions throughout the text
as ``ions at temperature $T$'', where $T$ is the temperature of the buffer
gas in the corresponding cooling simulation.

\section{Results and discussion}

\begin{figure}
\centering
    \includegraphics[width=\linewidth]{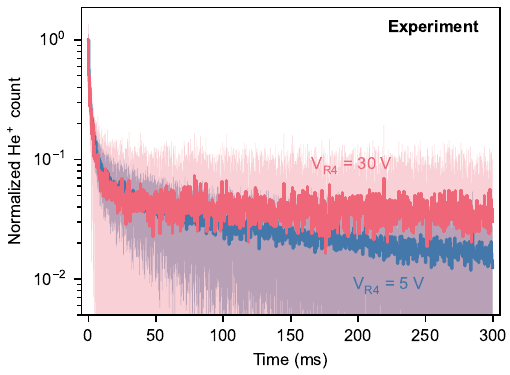}
    \caption{\label{fig:exp_tofa}
    Time of arrival of \Hep\ ions at the detector for a trap
    temperature of 36~K and two different voltages of the R4 ring
    electrode. Shaded areas represent the 
    uncertainty in the measurement.
    Values are normalized to the maximum for comparison.}
\end{figure}

The experimental data is obtained as time-resolved histograms of
the ions arriving at the detector. Two examples of this are given in
Fig.~\ref{fig:exp_tofa}. Typically, a large fraction of
the ions arrive within a few ms of
the trap opening (switching the SA voltage from $+0.3$ to $-1.5$~V),
and the signal decays sharply and
stabilizes after $\sim 10$~ms
to a value between $\sim 1-10\%$, depending on the particular \VRe\ voltage
and trap temperature.
From that point on to the end of the measurement
at 300~ms, a roughly constant countrate is observed. An exception to this is
found at very low \VRe\ voltages, where the trap is significantly depleted over
the 300~ms measurement and therefore an exponential decay trend is observed for
the longer timescales, as illustrated by the $\VRe = 5$~V measurement in that
figure. These two markedly different behaviors at short and
long timescales hint at two
different mechanisms through which the ions can escape the trap. For this reason,
we have chosen to process the experimental results in two ways, by integrating
the time-resolved measurements in full or only for the first ms.

\begin{figure}
\centering
    \includegraphics[width=\linewidth]{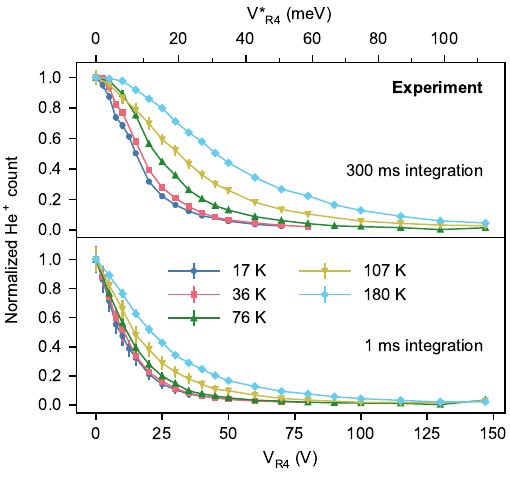}
    \caption{\label{fig:He_exp} \Hep\ ions extracted as a function of the
    voltage applied to the R4 ring electrode for different trap temperatures.
    Upper panel: 300~ms integration time. Lower panel: 1~ms integration time.
Top axis shows the calculated potential barrier for each R4 voltage. Ion count
is normalized to the maximum for each curve, which corresponds to $\VRe = 0$.}
\end{figure}

The experimental results of the energy sampling measurements are shown in
Fig.~\ref{fig:He_exp}.  The ion signal after opening the exit electrode was
recorded for different trap temperatures and ring electrode voltages and then
integrated in the aforementioned two ways.

The 300~ms integration time results in a slow decay of the ion signal as the
ring electrode voltage is increased. A small plateau in the very first \VRe\
voltages (up to $\sim 10$~V for the 180~K data)
is followed by a decay that is dependent on the ion temperature,
although the data for 17 and 36~K present very similar behaviors.
Full trapping can be achieved at $\VRe = 75$~V for the two
lowest temperatures.
At the maximum available ring electrode voltage, 147~V, ions are
effectively trapped for all except
the 180~K experiment, in which
$\sim 5\%$ of the ions are still able
to escape over the barrier.

Reducing the integration time to 1~ms results in a notable change in the shape
of the curves. A steady drop is now present from the beginning, and the signal
decays faster with increasing ring electrode voltage. On the other hand, much
smaller differences are observed for the curves at different temperatures,
particularly for the data of 76~K and below.
Even with the faster decay, the ions are still effectively trapped at similar
voltages as in the 300~ms integration time.

\begin{figure*}
\centering
    \includegraphics[width=\linewidth]{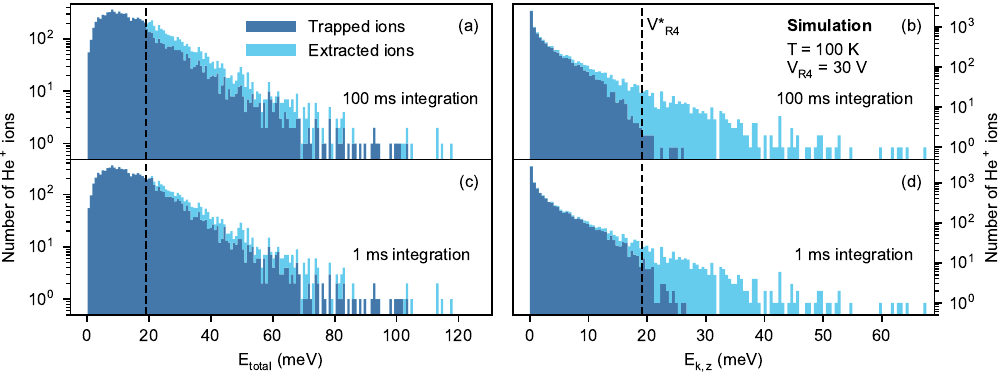}
    \caption{\label{fig:energies_cutoff}
    (Color online) Distribution of trapped (blue) and
    extracted (cyan, stacked on top) ions according to the initial values of
    their total (panels (a) and (c)) and axial kinetic (panels (b) and (c))
    energies, for 10000 ions at 100~K and $\VRe = 30$~V. Dashed black lines
    indicate the value of the the mechanical potential barrier. Top panels:
    results after the full 100~ms simulation. Bottom panels: results after
1~ms.} \end{figure*}

To better understand the way ions are behaving inside the trap and the exact
conditions that allow them to escape, we can look at the results from the ion
trajectory simulations. The first question to address is the correlation
between the energy of the ions and their likelihood to escape from the trap.
Figure~\ref{fig:energies_cutoff} depicts the total energy (\Etotal, sum of
kinetic and potential) and axial kinetic energy (\Ekz) distribution of a
simulation of ions at 100~K, showing which ones are able to escape the trap. As
expected, no ion with a total energy below the barrier value is extracted, but
on the other hand ions with energies 3--4
times higher than the ring electrode
barrier remain trapped. Conversely, looking at the axial kinetic energy
distribution, ions with \Ekz\ larger than the barrier are almost guaranteed to
exit the trap. However, in the full 100~ms simulation, a significant fraction
of the ions with \Ekz\ below the barrier are also able to escape. A better
correlation can be obtained by only considering the results after the first ms
of the simulation. In this case, most of the ions with axial kinetic energy
below the barrier remain trapped, while most of the higher energy ones are
still extracted.

\begin{figure}
\centering
    \includegraphics[width=\linewidth]{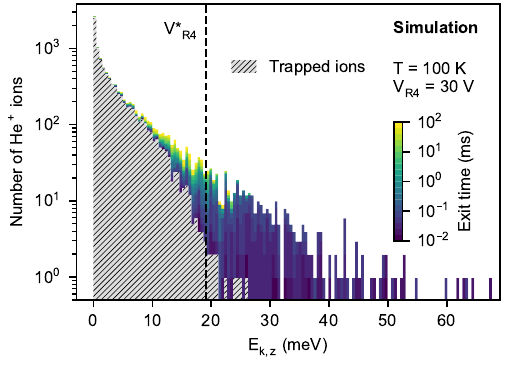}
    \caption{\label{fig:enz_time}
    (Color online) Initial axial kinetic energy distribution for
    the same conditions as Fig.~\ref{fig:energies_cutoff}. Extracted ions are
    now color-coded as a function of the time needed to exit the trap.
    The hatch pattern represents the fraction of ions
    that do not escape the trap.}
\end{figure}

A more detailed look at the time dependence of the extraction is shown in
Fig.~\ref{fig:enz_time}. Here, the different colors for the extracted ions
correspond to the time elapsed from the beginning of the simulation until they
reach the exit electrode SA.  As can be seen, most of the ions with initial
axial kinetic energies higher than the barrier need very little time
($\lesssim$ 0.1~ms) to exit the trap.
At energies lower than \VeffRe, times grow to
the scale of ms and a significant fraction (note the logarithmic scale)
of the ions require
close to the full simulation time to be able to escape.
As a reference, $\sim$10\% of all ions that leave the trap do so after 50~ms.
The fact that ions with
$\Ek < \VeffRe$ can still escape the trap is not unexpected, as their initial
potential energy might be enough to accelerate them to the needed velocity, but
the long times required by some of them to reach the exit
point towards an inefficient exchange of energy between the axial and
radial/angular degrees of freedom causing the axial energy to become
higher than the \VeffRe\ barrier at a late point in the simulation. A more
detailed look into the particulars of the ion motion inside the trap and
how this affects the energy exchange can be found in Appendix
\ref{app:ion_motion}.

\begin{figure}
\centering
    \includegraphics[width=\linewidth]{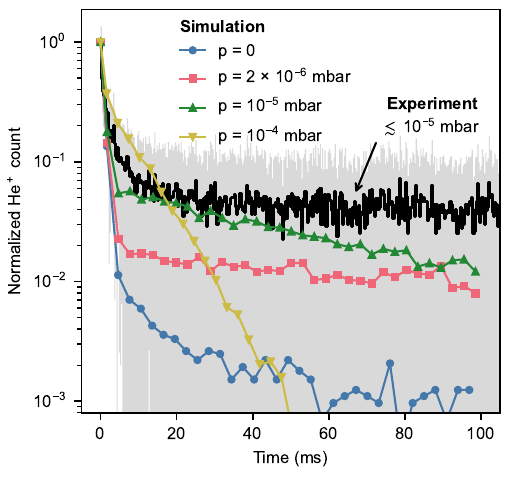}
    \caption{\label{fig:sim_tofa}
    (Color online) Experimental (bold black line, with 
    uncertainty
    shaded in gray) and simulated
    (color lines with symbols) histograms for the
    extraction times of \Hep\ ions with a \VRe\ voltage of 30~V. Experiment:
    time of arrival of the ions at the detector for a trap temperature of
    107~K.  Simulations: time elapsed from the beginning of the simulation
    until the ions reach the exit electrode, for ions at 100~K, with different
    residual buffer gas pressures. Values are normalized to the maximum for
comparison.}
\end{figure}

The inefficient coupling of the different degrees of freedom effectively
results in two different regimes when analyzing the energy sampling, separating
the first few ms of sampling from the rest of the acquisition.
As was shown in Fig.~\ref{fig:exp_tofa}, this is also apparent in
the time-resolved detection of the extracted
ions. A comparison of the experimental data with the simulated time-resolved
extraction of the ions is shown in Fig.~\ref{fig:sim_tofa}.
In both cases, a large fraction of the
ions is able to escape the trap in the first few ms. The experimental signal
then decays smoothly and is mostly stabilized to a value around 5\% of the
initial count.
In contrast, the simulations predict a sharper decay after the
initial extraction. Very few ions (a fraction 20 times smaller
than the experimental value)
escape after $\sim 20$~ms in the simulations
with no buffer gas, due to the aforementioned inefficient coupling between the
axial and radial motion of the ions. The extraction in this region can however
be significantly enhanced by including a small pressure of He buffer gas
in the simulations (which might
remain in the trap from the initial He pulse
in the experiment).
Collisions with this residual gas effectively repopulate the high
energy tail of the distribution, allowing more ions to overcome the potential
barrier. This leads to a factor
of $\sim 10$ increase in the extraction at 100~ms
with just $2 \times 10^{-6}$~mbar of residual buffer gas in the trap. Further
increasing the pressure in the simulations leads to increased extraction,
such that
a significant fraction of the trapped ions are able to escape, leading to the
depletion of the trap and
an exponential decay of the \ce{He+} count with time. This can be clearly
observed in the $10^{-4}$~mbar simulation.

As mentioned in the experimental section, the time evolution of the pressure
inside the trap during the experiments
is not reliably resolved by the pressure gauges.
Nevertheless, the manometer readings suggest that similar pressures are
attained at the different trap temperatures of this study. For this reason,
we use the residual gas
pressure instead of number density as a parameter for the trajectory
simulations.
For illustration, the value used for most of the simulations presented
here, $2 \times 10^{-6}$~mbar, corresponds to $8.4 \times 10^{10}$~\densU\ at
100~K and $2.2 \times 10^{11}$~\densU\ at 15~K.

Residual buffer gas can then help explain the observed high extraction at
longer timescales, and could also be a reason for the experimentally observed
smooth decay at the beginning due to the evacuation of this gas from the trap
with time, which is not accounted for in the simulations. 
Nevertheless, a
discrepancy still exists between the pressure needed to account for the signal
at longer extraction times and the one that better reproduces the experimental
results from Fig.~\ref{fig:He_exp} (see also Fig.~\ref{fig:sim_extraction}
below).

\begin{figure*}
\centering
    \includegraphics[width=\linewidth]{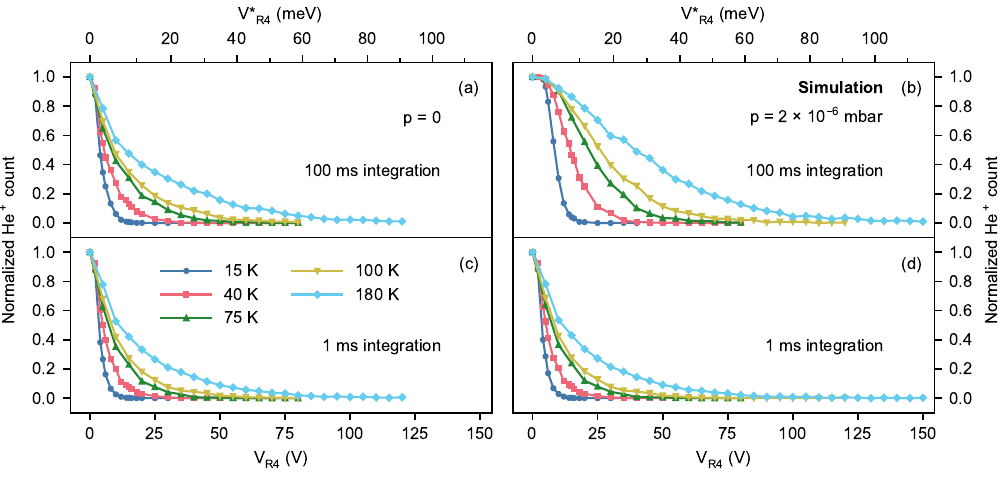}
    \caption{\label{fig:sim_extraction} Simulated number of extracted ions as a function of \VRe\ potential for different trap temperatures, with no residual buffer gas in the trap
    (panels (a) and (c)) and with a buffer gas pressure of $2 \times
10^{-6}$~mbar (panels (b) and (d)). Upper panels: fraction of ions extracted in
the full 100~ms simulation. Lower panels: fraction of ions that escape the trap
in the first ms.} \end{figure*}

Panels (a) and (c) of Fig.~\ref{fig:sim_extraction} show the simulated energy
sampling experiment for ions at different temperatures, with no residual buffer
gas present in the trap. As in Fig.~\ref{fig:He_exp}, the two panels show the
long-time integrated extraction and the first ms signal respectively. In the
case of the simulations, there is only a minor difference between both
processing methods, due to the low amount of ions leaving the trap after the
initial few ms as seen above. Qualitatively, the extraction curves for the 1~ms
integration resemble their experimental counterparts the most, with a sharply
decreasing trend for the lower \VRe\ voltages. On the other hand, the
experimental curves show a more extended profile, and a smaller separation
between the different temperatures, especially below 100~K.

The stark difference between simulation and experiment for the 100~ms
integration can be somewhat mitigated by including the effect of buffer gas
collisions in the sampling. Panels (b) and (d) of Fig.~\ref{fig:sim_extraction}
depict again both integration intervals, but now with a residual gas pressure
of $2 \times 10^{-6}$~mbar. The curves for 100~ms and 1~ms are now well
differentiated. In the 100~ms case, the enhanced extraction at longer
timescales observed in Fig.~\ref{fig:sim_tofa} translates into smoother curves
presenting a plateau for the first few \VRe\ values. This behavior was also
observed in the experiments, particularly for the 180~K curve. For higher ring
electrode voltages, the extraction, particularly at higher temperatures, is
significantly enhanced by the presence of buffer gas.
Limiting the sampling to the first ms, however, results in a figure almost
identical to the one obtained with no buffer gas in the trap, since buffer gas
collisions mainly contribute to the extraction at
timescales $\gtrsim 5$~ms.

It should be noted that the value of $2 \times 10^{-6}$~mbar is only chosen as
a reasonable pressure that results in relatively good agreement with the
sampling experiments. If the presence of residual gas from the cooling He pulse
is actually the cause of the observed experimental behavior, then the pressure
inside the trap will change over the extraction interval, and its value will
very likely be different for the various trap temperatures studied. Thus, the
simulated curves (with 100 ms integration) should only be taken as a
qualitative example of the type of changes produced by the presence of residual
buffer gas in the trap.
Increasing the trapping time, \ie, the time between the He
buffer gas pulse and the opening of the exit electrode, would reduce the
effect of the residual gas in exchange for an increased measurement time.
However, potential applications of this technique, such as the study of
exothermic reactions inside the trap (see Fig.~\ref{fig:H3p} below),
require a certain amount of neutral
gas to be present in the trap, hence, the effect of neutral gas collisions
cannot be completely avoided.

\begin{figure}
\centering
    \includegraphics[width=\linewidth]{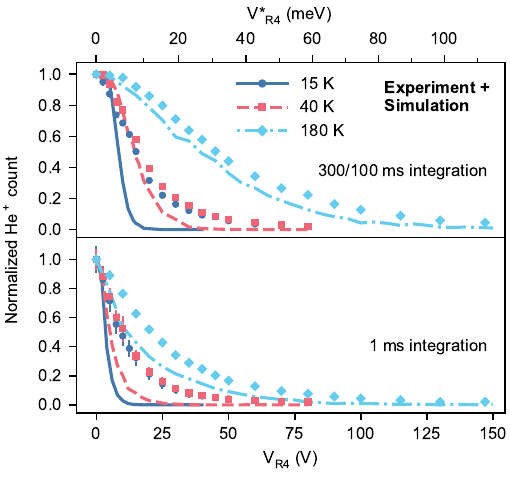}
    \caption{\label{fig:He_exp_sim}
    Comparison of experimental and simulated number of extracted ions as a
    function of \VRe\ voltage for different trap temperatures.
    Symbols represent experimental data, lines correspond to simulation results.
    The experimental trap temperature values are 17, 36 and 180~K respectively
    (only three sets of data are plotted for clarity).
    Upper panel: long integration time (300~ms for experimental data,
    100~ms for simulations). Lower panel: 1~ms integration time.
    Simulation results correspond to a residual gas pressure of
    $2 \times 10^{-6}$~mbar (see text).
    }
\end{figure}

A direct comparison of the experimental and simulated extraction curves
for three different temperatures can be found in
Fig.~\ref{fig:He_exp_sim}. For the simulations, the results with a
residual pressure have been used for the comparison in both panels,
since it results in a
relatively good agreement for the data with long integration times
(300 or 100~ms for experiment and simulation,
respectively),
while for the 1~ms integration times, the simulations with and
without residual gas pressure result in virtually the same curve.
For long integration times, the simulation results show a reasonable agreement
with the experiments, particularly at the higher temperatures.
At 180~K, the simulated curve follows the
experimental data quite closely, but always results in an
underestimation of about 10\%
of the fraction of extracted ions. The agreement worsens
at lower temperatures. At 40~K, the simulations deviate from the
experiments for $\VRe > 20$~V, and the 15~K simulation significantly
falls significantly below the experimental results. The discrepancy increases
for the data with 1~ms integration time, where the simulations always
predict much narrower curves than what is observed experimentally. Further
details on the differences between the experimental and
simulated results are presented in Appendix \ref{app:energy_sampling}.
The experimentally enhanced extraction of ions for high \VRe\ voltages
was also observed by \citet{lakhmanskayaPropertiesMultipoleIon2014} in their
evaporation study in a 22-pole trap with constant buffer gas pressure.
In their case, the temperature
derived from this enhanced loss rate resulted in almost a factor of 2 increase
with respect to the experimental trap temperature.

These differences between the
measurements and the expected results from the simulation are significant
enough to prevent the derivation of the energy distribution from the extraction
data. There are multiple potential causes for this disagreement. First and
foremost is the possible presence of patch potentials and stray fields due to
manufacturing imperfections changing the extraction process, either through the
distortion of the ring electrode barrier or simply by affecting the motion or
energy of the trapped ions.
Limitations in the simulation, such as the usage
of the ideal multipole expression for the rf field, could also play a role.
The inefficient motion coupling derived from the
simulations seems to be reinforced by the different experimental results at 1
and 300~ms integration, but a more effective coupling may also account for
these together with the MCS time-resolved measurements, instead of, or together
with the proposed effect of residual buffer gas pressure.
An increased efficiency of the motion coupling of the ions was also proposed
by \citet{notzoldSpectroscopyIonThermometry2022} to explain the
comparable effect
of rf heating measured along the axial and radial directions of their wire
trap. We have also assumed through this work that the increased ring electrode
voltage has no effect on the experiment other than the increase of the
potential barrier in the trap. However, it is possible that the focusing of the
ions escaping the trap is dependent on the {\VRe} potential in the subsequent
ion optics, affecting their overall detection efficiency.
Ultimately, the use
of this technique for thermometry measurements requires, effectively, an
accurate knowledge of the potential inside the trap, especially the ring
electrode barrier, and, potentially, an experimental
calibration of this barrier
using ions of known energy might be a better suited approach to
refine the energy sampling method. 
This may be accomplished by launching ions directly from the
SIS into the trap and stopping them with either the ring or exit electrodes,
providing a relation between both potentials. Another possibility is to
use a process with well known energetics, such as the quenching of an
excited ion state in a LOS experiment, to calibrate the barrier.

Despite these caveats, the proposed energy sampling setup is still proven to
reliably distinguish between ion distributions with characteristic energies
that differ by tens of meV. Furthermore, the effect of the collisions with
neutral gas is found to be confined to long integration times, meaning that, in
principle, the results of this sampling method should be consistent even with
some pressure present in the trap, as long as the characteristic collisional
time is larger than the short-scale sampling window.  This opens up the
possibility of using this method to study a variety of ion-molecule processes
in the trap, such as the energetics of exothermic ion-molecule reactions.

An example of one such measurement is shown in
Fig.~\ref{fig:H3p}. Energy sampling experiments were conducted for
\HHHp\ ions produced in different ways and at different trap temperatures.
Ions that were produced in the Storage Ion Source (SIS)
(in the same way as the \Hep\ presented above) present extraction
curves depending on their temperature, with a behavior that is quite
similar to the \Hep\ experiments. A markedly different result is obtained
when the \HHHp\ ions are produced inside the trap. In this case,
\ce{H2+} ions are stored in the trap together with
a very low number density
($\sim 1 \times 10^{9}$~\densU) of neutral \ce{H2} gas,
so that \HHHp\ is produced through the 1.7~eV exothermic reaction
\ce{H2+ + H2 -> H3+ + H}. A fraction of the energy released ends up as
translational energy of the \HHHp\ ion, which is clearly visible in the
experiment, as the ring electrode voltages needed to contain the ions inside
the trap are $\sim$ 3--4 times larger than for thermal ions. Secondary
processes, such as the quenching of the vibrationally excited ions produced
in the reaction, will also contribute to this increased high kinetic energy
tail. This makes the extraction of precise information about
the energetics of the reaction quite challenging and will be the
subject of future work.
Furthermore, collisions with the neutral gas
are a necessary part of the experimental procedure, and, as seen before,
these collisions hinder the recovery of the energy distribution information
from the extraction curve. Nevertheless, the very clear effect of the
exothermic process on the experimental results suggests that the distinction
of ions of the same mass produced in the trap, as is
the case with reactions producing different isomers
(with different exothermicity, \ie, kinetic energies), can be
achieved with this technique.
Although we clearly demonstrate that our method is sensitive even below
40~K for \Hep, we expect further experimental optimization will be required in
order to study processes involving $<4$ \mz\ ions, like the proton
exchange $\ce{H+} + \ce{H2}(o) \to \ce{H+} + \ce{H2}(p)$ leading to the
\emph{ortho} to \emph{para} conversion of the hydrogen molecule
\cite{gerlichOrthoParaTransitions1990},
with exoergicity as low as $<15$~meV ($<175$~K).

\begin{figure}
\centering
    \includegraphics[width=\linewidth]{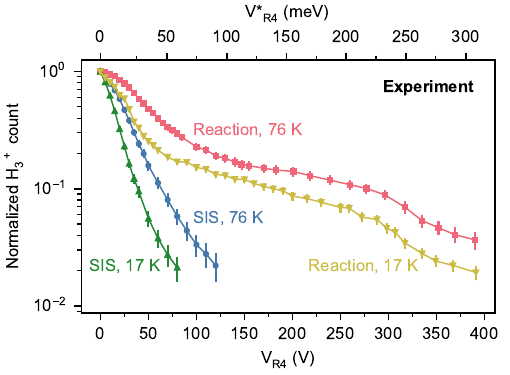}
    \caption{\label{fig:H3p}
    (Color online) Experimental number of \HHHp\
    ions extracted as a function of the \VRe\
    voltage for two different trap temperatures and production methods.
    Blue circles and green upward triangles: \HHHp\ ions produced in the Storage
    Ion Source (SIS) and subsequently trapped. Red squares and yellow
    downward triangles: \HHHp\ ions created in the trap through the exothermic
    reaction \ce{H2+ + H2 -> H3+ + H}. Integration times are 300~ms for
    all measurements.
}
\end{figure}

\section{Summary and conclusions}

We present an experimental method for the characterization of ion energies in a
22-pole trap using ring electrodes. By inducing small potential barriers before
the trap exit, the trapped ions can be selectively extracted depending on their
energy and subsequently detected. The experimental investigation of this method
has been complemented by numerical simulations in order to identify the role of
various processes, such as ion motion coupling or residual buffer gas
collisions, in the expected results.

\Hep\ ions have been confined at different trap temperatures and subsequently
sampled in order to test the experimental method, obtaining different curves
for the extracted ion fraction as a function of the applied ring electrode
voltage. These curves, which can be likened to energy distributions, are
observed to get narrower (\ie, trend towards lower energies) as the trap
temperature is decreased. 
Comparison of the experimental and simulated time-resolved ion detection
suggests that residual buffer gas in the trap may play a role
in the extraction process. 
The energy sampling measurements show
discrepancies with the simulations
that may be caused by differences between the simulated and real
trap potentials or a more efficient motion coupling of the ions than the one
predicted in the model. 
While these discrepancies 
prevent the derivation of the energy distribution from the extraction data,
an experimental calibration of the potential barrier may suffice to refine
the energy sampling method.

The method as presented is nevertheless proven to reliably
discriminate between ion distributions with characteristic energies that differ
by tens of meV, and thus,
we expect this energy sampling method to be useful in the
study of the energetics of ion-molecule processes in the ion trap, as well as
in distinguishing ions in situations where mass spectrometry is not sufficient,
such as when dealing with isomers or internally excited species.
Another prime target of this method are studies of the
energy transfer of internal excitation to translational
energy through inelastic collisions, which can consequently
directly lead to optimization of the LOS setups
\cite{schmidLeakOutSpectroscopyUniversal2022,
Bast2023}. The possibility to modify the ion energy
distribution in the trap by selective evaporation
puts this technique ahead of other
sampling methods relying on either full ejection of the ion cloud
\cite{champeauKineticEnergyMeasurements1994} or on buffer gas induced
evaporation \cite{lakhmanskayaPropertiesMultipoleIon2014}.
We have already showcased the usefulness of this
method in highly exothermic ion-molecule reactions, by distinguishing ions
created in the trap from those produced in the Storage Ion Source.

The ion trajectory code employed in this work can be readily extended to
include reactive collisions in which the nascent ions gain high kinetic
energies, either from the exothemicity of the reaction, or from internal
degrees of freedom, \eg, induced by absorbed photon, supporting our future
experimental work focused on processes
between ions and neutrals at astrophysically relevant conditions.
We intend to explore the possible broadening of the method towards mass
spectrometry inside the trap, \ie, trap and ring electrode configuration
in which ions are selectively ejected according to their \mz, first
computationally (code extensions), then followed by the experiment.

\subsection*{Data Availability} The supporting data for this article are openly
available from the zenodo archive DOI:
\href{https://doi.org/10.5281/zenodo.8321098}{10.5281/zenodo.8321098}.

\begin{acknowledgments}

    This work was supported by the Max Planck Society.  The authors gratefully
    acknowledge the work of the electrical and mechanical workshops and
    engineering departments of the Max Planck Institute for Extraterrestrial
    Physics.  We thank Prof. Stephan Schlemmer for helpful discussions.  We
    thank the Lorentz Center, Leiden, the Netherlands for organising the ``New
    Directions in Cold and Ultracold Chemistry'' workshop and its participants
    for fruitful discussions.

\end{acknowledgments}

\appendix

\section{Coupling of the ion motion} \label{app:ion_motion}

 The ion motion is dominated by the effective potential $V^*$ (see
Fig.~{\ref{fig:Veff}}), which governs the macromotion of the ions.
Accordingly, the possible coupling of the different
degrees of freedom will be determined by the interaction of the ions
with this potential.

The effective potential created by an ideal multipole
of order $n$ as a function of the radial coordinate $r$
scales as $V^* \propto r^{2n-2}$ (see for instance
\cite{gerlichInhomogeneousRFFields1992}). As a consequence,
a significant fraction of a 22-pole trap volume is virtually a
field free region, which is one of the advantages of this design.
Close to the poles, however, this potential grows very rapidly. As a
consequence, ions moving towards the poles are essentially reflected
by this effective potential and should retain their energy distribution
across the different degrees of freedom.

In a real trap, however, the end electrodes and ring
electrodes will also contribute to the effective potential. In the
central region, sufficiently far from the ends, their contribution is
small (see Fig.~\ref{fig:Veff} and \cite{juskoColdCASIon2023})
and the assumptions of the previous
paragraph for radial motion should hold. The situation gets more
complicated for the axial motion. The potential does not grow as
fast in that direction compared to the radial one, and there is a weak
dependence also along the radial direction for a given axial
coordinate. It can then be expected that, as the ions change trajectory
due to this potential, some of the deflections result in a transfer of
kinetic energy between the different degrees of freedom. This effect
should be particularly relevant if the ion moves towards the corners
of the trap, where the contributions of both end electrodes and poles
are significant. How often this transfer of energy happens will likely
depend on the particular ion motion and the electrode configuration
and can be better examined in our simulations.

\begin{figure}
\centering
    \includegraphics[width=\linewidth]{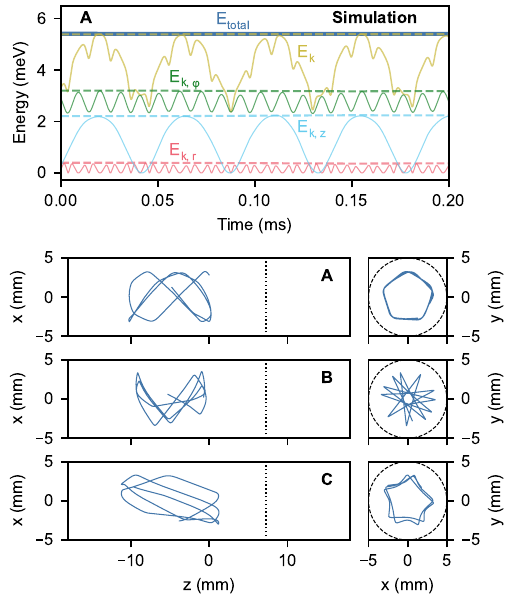}
    \caption{\label{fig:traj_energy}
    (Color online) Upper panel: simulated time evolution of the
    total kinetic energy ($E_k$, yellow) and the
    kinetic energy for the different degrees
    of freedom, radial ($E_{k,r}$, red),
    angular ($E_{k,\varphi}$, green)
    and axial ($E_{k,z}$, cyan), along with
    the total energy (sum of kinetic and potential, bold blue line),
    for a trapped
    {\Hep} ion from a 15~K distribution residing in the trap with a
    $\VRe = 30$~V barrier and no buffer gas collisions,
    during the first 0.2~ms of the
    simulation (time A). Dashed lines represent
    an envelope obtained from the
    maximum values
    of the different energies
    in 0.1~ms bins. Lower panels: projection of the ion
    trajectory during 0.2~ms in the $xz$ and $xy$ planes,
    for the aforementioned ion at three different times of the 100~ms
    simulation, A, B, C,
    corresponding to motion dominated by angular,
    radial, and axial velocities respectively (see
    Fig.~{\ref{fig:energy_degrees}} for details). The vertical dotted line
    represents the position of the ring electrode barrier, the circular
    dashed line shows the inscribed radius $r_0$ of the 22 rods.}

    \end{figure}

An example of the motion of a single trapped ion
moving according to the effective potential of the trap is shown in
Fig.~{\ref{fig:traj_energy}}.
The time evolution of the kinetic energy ({\Ek}) and the contribution from
the different degrees of freedom
(radial, $r$, angular, $\varphi$, and axial, $z$) in
a sub-ms timescale is shown in the upper panel and its trajectory in the
$xz$, $xy$ planes in panel A.
As the ion moves inside the trap, it bounces back and forth due to the
confining field, producing the oscillatory behavior observed for the
kinetic energy.
These oscillations are nevertheless found to have regular amplitudes
as seen from the almost constant envelopes obtained from the respective
maxima, meaning that the ion is not exchanging kinetic energy between the
different degrees of freedom at $\ll \text{ms}$ timescales.
Several distinctive motion patterns, with distinctive \Ek\ components,
have been identified (A--C).
A particular ion may change the motion pattern,
even without undergoing a collision with a neutral.

\begin{figure}
\centering
    \includegraphics[width=\linewidth]{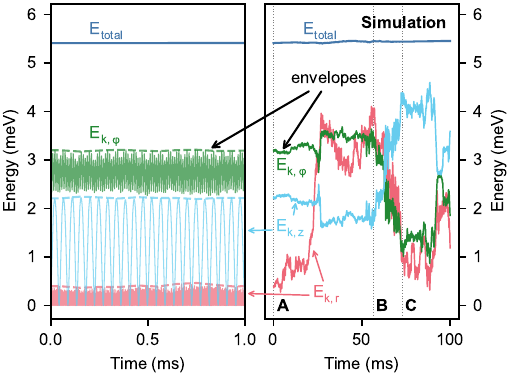}
    \caption{\label{fig:energy_degrees} Simulated
    coupling of the kinetic energy
    in the different degrees
    of freedom,
    radial ($E_{k,r}$, red),
    angular ($E_{k,\varphi}$, green)
    and axial ($E_{k,z}$, cyan), along with
    the total energy (sum of kinetic and potential, bold blue line)
    for the same ion trajectory depicted in
    Fig.~\ref{fig:traj_energy}.
    Left panel: time evolution
    during the first ms (solid lines), along with the envelopes
    obtained from the
    maximum values (dashed lines). Right panel: time evolution
    for the full 100~ms simulation,
    using the aforementioned envelopes for clarity.
    The dotted vertical lines show the times A, B and
    C corresponding to the trajectory plots in
    Fig.~\ref{fig:traj_energy}.
    Note that there are no collisions, \ie, the ion motion, and the coupling
    between the \Ek\ components, is governed solely by the effective
    potential.}

    \end{figure}

Since the axial kinetic energy controls whether the ion is able to escape
the trap, the potential
coupling of the energy in the different degrees of freedom for the motion of
the ion will play a significant role in the sampling process. This is
examined in Fig.~{\ref{fig:energy_degrees}}. The left
panel expands the results of Fig.~{\ref{fig:traj_energy}} to a 1~ms interval,
and still, no significant exchange between the different kinetic
energy contributions can be observed, \ie, within this
time frame only the initial axial velocity and potential energy will play a
role in the extraction of the ion. For longer timescales, however, the
situation changes dramatically,
as seen in the right panel of Fig.~\ref{fig:energy_degrees}.
Along the 100~ms simulation, the maximum kinetic energy in the axial direction
varies significantly, accounting for
a minimum of 1.6~meV to a maximum of 4.6~meV of the
total ion energy of 5.4~meV. This implies that the coupling between different
\Ek\ components is possible in this timescale
and an ion can escape the trap
long after the start of the simulation
if its axial kinetic energy becomes higher than the
\VeffRe\ potential barrier.

For a complete description of this coupling, the effect of the
micromotion caused by the rf field should also be considered.
The analysis of the ion kinetic energy,
however, becomes much more complex in this case, as the ions gain large amounts
of energy when moving closer to the rods where the rf field is stronger.
This means that it is no longer possible to examine the coupling of the motion
by simply looking at the kinetic energy envelopes as in
Fig.~\ref{fig:energy_degrees}.

Nevertheless, the behavior observed for the axial energy \Ekz, which is not
affected by this micromotion, is similar to the one in the effective potential
simulations.
This fact, coupled with the similar
results of the energy sampling simulations performed
using the effective potential and the time-dependent rf field, suggests
that the micromotion is not a major factor in the coupling of the different
degrees of freedom.

\section{Detailed energy sampling analysis} \label{app:energy_sampling}
The experimental and simulation results for the energy sampling
can be further analyzed by
using the value of the \VRe\ voltage at which
the fraction of extracted ions is 0.5 (\Vhalf) and 0.1 (\Vtenth).
These are collected in Table~\ref{tab:trapping_voltages}, including those
for the 75 and 100~K data omitted in Fig.~\ref{fig:He_exp_sim}.
For the long integration times, the values of \Vhalf\ are reasonably well
reproduced by the simulations, except at 15~K, but the \Vtenth\ values are
consistently underestimated, meaning the higher barrier values can be
overcome by a larger amount of ions than expected from the simulations.
At 1~ms the discrepancy is much larger, with the experimental \Vhalf\ values
almost doubling those obtained from the simulations, and \Vtenth\ voltages
that diverge particularly at the lower temperatures. Despite the 1~ms
integration time largely sidestepping the effect of collisions with the
neutral gas, it is unclear whether that introduces additional issues in
the sampling processes leading to the observed discrepancies.

\begin{table*}
\caption{Voltages at which the fraction of ions extracted falls to 0.5 (\Vhalf)
and 0.1 (\Vtenth) for the experimental (Exp) and simulation (Sim) data
presented in Figs.~\ref{fig:He_exp}, \ref{fig:sim_extraction} and
\ref{fig:He_exp_sim}, depending on the integration time.
Simulation results correspond to a residual gas pressure of
$2 \times 10^{-6}$~mbar (see text).
}

\label{tab:trapping_voltages}
\begin{ruledtabular}
\begin{tabular}{c c || c c | c c || c c | c c}

& & \multicolumn{4}{c||}{300/100~ms} & \multicolumn{4}{c}{1~ms} \\
\hline
\multicolumn{2}{c||}{T~(K)} & \multicolumn{2}{c|}{\Vhalf~(V)}
& \multicolumn{2}{c||}{\Vtenth~(V)} & \multicolumn{2}{c|}{\Vhalf~(V)}
& \multicolumn{2}{c}{\Vtenth~(V)}\\
\hline
Exp & Sim & Exp & Sim & Exp & Sim & Exp & Sim & Exp & Sim \\
17 & 15 & $15.1 \pm 0.4$ & 8.5 & $28.9 \pm 1.4$ & 12.9 &
$9.2 \pm 1.9$ & 3.5 & $30.8 \pm 5.1$ & 7.3\\
36 & 40 & $17.1 \pm 0.4$ & 14.6 & $42.0 \pm 1.2$ & 26.2 &
$10.6 \pm 2.4$ & 5.2 & $32.2 \pm 3.6$ & 13.1\\
76 & 75 & $22.7 \pm 0.8$ & 21.7 & $56.9 \pm 1.9$ & 40.1 &
$11.8 \pm 1.2$ & 7.5 & $34.7 \pm 3.2$ & 22.2\\
107 & 100 & $31.3\pm 2.0$ & 26.0 & $81.7 \pm 6.0$ & 52.2 &
$14.4 \pm 2.2$ & 8.5 & $49.0 \pm 8.5$ & 27.5\\
180 & 180 & $45.4 \pm 1.2$ & 39.3 & $110.8 \pm 2.7$ & 85.5 &
$21.0 \pm 1.5$ & 11.6 & $68.4 \pm 4.7$ & 47.9\\
\end{tabular}
\end{ruledtabular}
\end{table*}


\begin{thebibliography}{34}%
\makeatletter
\providecommand \@ifxundefined [1]{%
 \@ifx{#1\undefined}
}%
\providecommand \@ifnum [1]{%
 \ifnum #1\expandafter \@firstoftwo
 \else \expandafter \@secondoftwo
 \fi
}%
\providecommand \@ifx [1]{%
 \ifx #1\expandafter \@firstoftwo
 \else \expandafter \@secondoftwo
 \fi
}%
\providecommand \natexlab [1]{#1}%
\providecommand \enquote  [1]{``#1''}%
\providecommand \bibnamefont  [1]{#1}%
\providecommand \bibfnamefont [1]{#1}%
\providecommand \citenamefont [1]{#1}%
\providecommand \href@noop [0]{\@secondoftwo}%
\providecommand \href [0]{\begingroup \@sanitize@url \@href}%
\providecommand \@href[1]{\@@startlink{#1}\@@href}%
\providecommand \@@href[1]{\endgroup#1\@@endlink}%
\providecommand \@sanitize@url [0]{\catcode `\\12\catcode `\$12\catcode `\&12\catcode `\#12\catcode `\^12\catcode `\_12\catcode `\%12\relax}%
\providecommand \@@startlink[1]{}%
\providecommand \@@endlink[0]{}%
\providecommand \url  [0]{\begingroup\@sanitize@url \@url }%
\providecommand \@url [1]{\endgroup\@href {#1}{\urlprefix }}%
\providecommand \urlprefix  [0]{URL }%
\providecommand \Eprint [0]{\href }%
\providecommand \doibase [0]{https://doi.org/}%
\providecommand \selectlanguage [0]{\@gobble}%
\providecommand \bibinfo  [0]{\@secondoftwo}%
\providecommand \bibfield  [0]{\@secondoftwo}%
\providecommand \translation [1]{[#1]}%
\providecommand \BibitemOpen [0]{}%
\providecommand \bibitemStop [0]{}%
\providecommand \bibitemNoStop [0]{.\EOS\space}%
\providecommand \EOS [0]{\spacefactor3000\relax}%
\providecommand \BibitemShut  [1]{\csname bibitem#1\endcsname}%
\let\auto@bib@innerbib\@empty
%</preamble>
\bibitem [{\citenamefont {N{\"o}tzold}\ \emph {et~al.}(2022)\citenamefont {N{\"o}tzold}, \citenamefont {Wild}, \citenamefont {Lochmann},\ and\ \citenamefont {Wester}}]{notzoldSpectroscopyIonThermometry2022}%
  \BibitemOpen
  \bibfield  {author} {\bibinfo {author} {\bibfnamefont {M.}~\bibnamefont {N{\"o}tzold}}, \bibinfo {author} {\bibfnamefont {R.}~\bibnamefont {Wild}}, \bibinfo {author} {\bibfnamefont {C.}~\bibnamefont {Lochmann}},\ and\ \bibinfo {author} {\bibfnamefont {R.}~\bibnamefont {Wester}},\ }\href {https://doi.org/10.1103/PhysRevA.106.023111} {\bibfield  {journal} {\bibinfo  {journal} {Physical Review A}\ }\textbf {\bibinfo {volume} {106}},\ \bibinfo {pages} {023111} (\bibinfo {year} {2022})}\BibitemShut {NoStop}%
\bibitem [{\citenamefont {Plašil}\ \emph {et~al.}(2023)\citenamefont {Plašil}, \citenamefont {Uvarova}, \citenamefont {Rednyk}, \citenamefont {Štěpán Roučka}, \citenamefont {Vanko}, \citenamefont {Dohnal},\ and\ \citenamefont {Glosík}}]{Plasil2023}%
  \BibitemOpen
  \bibfield  {author} {\bibinfo {author} {\bibfnamefont {R.}~\bibnamefont {Plašil}}, \bibinfo {author} {\bibfnamefont {L.}~\bibnamefont {Uvarova}}, \bibinfo {author} {\bibfnamefont {S.}~\bibnamefont {Rednyk}}, \bibinfo {author} {\bibnamefont {Štěpán Roučka}}, \bibinfo {author} {\bibfnamefont {E.}~\bibnamefont {Vanko}}, \bibinfo {author} {\bibfnamefont {P.}~\bibnamefont {Dohnal}},\ and\ \bibinfo {author} {\bibfnamefont {J.}~\bibnamefont {Glosík}},\ }\href {https://doi.org/10.3847/1538-4357/acc9ac} {\bibfield  {journal} {\bibinfo  {journal} {The Astrophysical Journal}\ }\textbf {\bibinfo {volume} {948}},\ \bibinfo {pages} {131} (\bibinfo {year} {2023})}\BibitemShut {NoStop}%
\bibitem [{\citenamefont {Endres}\ \emph {et~al.}(2017)\citenamefont {Endres}, \citenamefont {Egger}, \citenamefont {Lee}, \citenamefont {Lakhmanskaya}, \citenamefont {Simpson},\ and\ \citenamefont {Wester}}]{endresIncompleteRotationalCooling2017}%
  \BibitemOpen
  \bibfield  {author} {\bibinfo {author} {\bibfnamefont {E.~S.}\ \bibnamefont {Endres}}, \bibinfo {author} {\bibfnamefont {G.}~\bibnamefont {Egger}}, \bibinfo {author} {\bibfnamefont {S.}~\bibnamefont {Lee}}, \bibinfo {author} {\bibfnamefont {O.}~\bibnamefont {Lakhmanskaya}}, \bibinfo {author} {\bibfnamefont {M.}~\bibnamefont {Simpson}},\ and\ \bibinfo {author} {\bibfnamefont {R.}~\bibnamefont {Wester}},\ }\href {https://doi.org/10.1016/j.jms.2016.12.006} {\bibfield  {journal} {\bibinfo  {journal} {Journal of Molecular Spectroscopy}\ }\textbf {\bibinfo {volume} {332}},\ \bibinfo {pages} {134} (\bibinfo {year} {2017})}\BibitemShut {NoStop}%
\bibitem [{\citenamefont {Jusko}\ \emph {et~al.}(2014)\citenamefont {Jusko}, \citenamefont {Asvany}, \citenamefont {Wallerstein}, \citenamefont {Br\"unken},\ and\ \citenamefont {Schlemmer}}]{Jusko2014}%
  \BibitemOpen
  \bibfield  {author} {\bibinfo {author} {\bibfnamefont {P.}~\bibnamefont {Jusko}}, \bibinfo {author} {\bibfnamefont {O.}~\bibnamefont {Asvany}}, \bibinfo {author} {\bibfnamefont {A.-C.}\ \bibnamefont {Wallerstein}}, \bibinfo {author} {\bibfnamefont {S.}~\bibnamefont {Br\"unken}},\ and\ \bibinfo {author} {\bibfnamefont {S.}~\bibnamefont {Schlemmer}},\ }\href {https://doi.org/10.1103/PhysRevLett.112.253005} {\bibfield  {journal} {\bibinfo  {journal} {Phys. Rev. Lett.}\ }\textbf {\bibinfo {volume} {112}},\ \bibinfo {pages} {253005} (\bibinfo {year} {2014})}\BibitemShut {NoStop}%
\bibitem [{\citenamefont {Brünken}\ \emph {et~al.}(2017)\citenamefont {Brünken}, \citenamefont {Kluge}, \citenamefont {Stoffels}, \citenamefont {Pérez-Ríos},\ and\ \citenamefont {Schlemmer}}]{Brunken2017}%
  \BibitemOpen
  \bibfield  {author} {\bibinfo {author} {\bibfnamefont {S.}~\bibnamefont {Brünken}}, \bibinfo {author} {\bibfnamefont {L.}~\bibnamefont {Kluge}}, \bibinfo {author} {\bibfnamefont {A.}~\bibnamefont {Stoffels}}, \bibinfo {author} {\bibfnamefont {J.}~\bibnamefont {Pérez-Ríos}},\ and\ \bibinfo {author} {\bibfnamefont {S.}~\bibnamefont {Schlemmer}},\ }\href {https://doi.org/10.1016/j.jms.2016.10.018} {\bibfield  {journal} {\bibinfo  {journal} {Journal of Molecular Spectroscopy}\ }\textbf {\bibinfo {volume} {332}},\ \bibinfo {pages} {67} (\bibinfo {year} {2017})}\BibitemShut {NoStop}%
\bibitem [{\citenamefont {N{\"o}tzold}\ \emph {et~al.}(2020)\citenamefont {N{\"o}tzold}, \citenamefont {Hassan}, \citenamefont {Tauch}, \citenamefont {Endres}, \citenamefont {Wester},\ and\ \citenamefont {Weidem{\"u}ller}}]{notzoldThermometryMultipoleIon2020}%
  \BibitemOpen
  \bibfield  {author} {\bibinfo {author} {\bibfnamefont {M.}~\bibnamefont {N{\"o}tzold}}, \bibinfo {author} {\bibfnamefont {S.~Z.}\ \bibnamefont {Hassan}}, \bibinfo {author} {\bibfnamefont {J.}~\bibnamefont {Tauch}}, \bibinfo {author} {\bibfnamefont {E.}~\bibnamefont {Endres}}, \bibinfo {author} {\bibfnamefont {R.}~\bibnamefont {Wester}},\ and\ \bibinfo {author} {\bibfnamefont {M.}~\bibnamefont {Weidem{\"u}ller}},\ }\href {https://doi.org/10.3390/app10155264} {\bibfield  {journal} {\bibinfo  {journal} {Applied Sciences}\ }\textbf {\bibinfo {volume} {10}},\ \bibinfo {pages} {5264} (\bibinfo {year} {2020})}\BibitemShut {NoStop}%
\bibitem [{\citenamefont {Williams}\ \emph {et~al.}(2017)\citenamefont {Williams}, \citenamefont {Truppe}, \citenamefont {Hambach}, \citenamefont {Caldwell}, \citenamefont {Fitch}, \citenamefont {Hinds}, \citenamefont {Sauer},\ and\ \citenamefont {Tarbutt}}]{Williams2017}%
  \BibitemOpen
  \bibfield  {author} {\bibinfo {author} {\bibfnamefont {H.~J.}\ \bibnamefont {Williams}}, \bibinfo {author} {\bibfnamefont {S.}~\bibnamefont {Truppe}}, \bibinfo {author} {\bibfnamefont {M.}~\bibnamefont {Hambach}}, \bibinfo {author} {\bibfnamefont {L.}~\bibnamefont {Caldwell}}, \bibinfo {author} {\bibfnamefont {N.~J.}\ \bibnamefont {Fitch}}, \bibinfo {author} {\bibfnamefont {E.~A.}\ \bibnamefont {Hinds}}, \bibinfo {author} {\bibfnamefont {B.~E.}\ \bibnamefont {Sauer}},\ and\ \bibinfo {author} {\bibfnamefont {M.~R.}\ \bibnamefont {Tarbutt}},\ }\href {https://doi.org/10.1088/1367-2630/aa8e52} {\bibfield  {journal} {\bibinfo  {journal} {New Journal of Physics}\ }\textbf {\bibinfo {volume} {19}},\ \bibinfo {pages} {113035} (\bibinfo {year} {2017})}\BibitemShut {NoStop}%
\bibitem [{\citenamefont {Alt}\ \emph {et~al.}(2003)\citenamefont {Alt}, \citenamefont {Schrader}, \citenamefont {Kuhr}, \citenamefont {M\"uller}, \citenamefont {Gomer},\ and\ \citenamefont {Meschede}}]{Alt2003}%
  \BibitemOpen
  \bibfield  {author} {\bibinfo {author} {\bibfnamefont {W.}~\bibnamefont {Alt}}, \bibinfo {author} {\bibfnamefont {D.}~\bibnamefont {Schrader}}, \bibinfo {author} {\bibfnamefont {S.}~\bibnamefont {Kuhr}}, \bibinfo {author} {\bibfnamefont {M.}~\bibnamefont {M\"uller}}, \bibinfo {author} {\bibfnamefont {V.}~\bibnamefont {Gomer}},\ and\ \bibinfo {author} {\bibfnamefont {D.}~\bibnamefont {Meschede}},\ }\href {https://doi.org/10.1103/PhysRevA.67.033403} {\bibfield  {journal} {\bibinfo  {journal} {Phys. Rev. A}\ }\textbf {\bibinfo {volume} {67}},\ \bibinfo {pages} {033403} (\bibinfo {year} {2003})}\BibitemShut {NoStop}%
\bibitem [{\citenamefont {Hua}\ \emph {et~al.}(2019)\citenamefont {Hua}, \citenamefont {Feng}, \citenamefont {Zhou}, \citenamefont {Liang}, \citenamefont {Chen},\ and\ \citenamefont {Zhao}}]{huaCryogenicCylindricalIon2019}%
  \BibitemOpen
  \bibfield  {author} {\bibinfo {author} {\bibfnamefont {Z.}~\bibnamefont {Hua}}, \bibinfo {author} {\bibfnamefont {S.}~\bibnamefont {Feng}}, \bibinfo {author} {\bibfnamefont {Z.}~\bibnamefont {Zhou}}, \bibinfo {author} {\bibfnamefont {H.}~\bibnamefont {Liang}}, \bibinfo {author} {\bibfnamefont {Y.}~\bibnamefont {Chen}},\ and\ \bibinfo {author} {\bibfnamefont {D.}~\bibnamefont {Zhao}},\ }\href {https://doi.org/10.1063/1.5079264} {\bibfield  {journal} {\bibinfo  {journal} {Review of Scientific Instruments}\ }\textbf {\bibinfo {volume} {90}},\ \bibinfo {pages} {013101} (\bibinfo {year} {2019})}\BibitemShut {NoStop}%
\bibitem [{\citenamefont {Johnston}\ \emph {et~al.}(2018)\citenamefont {Johnston}, \citenamefont {Pearson}, \citenamefont {Wang},\ and\ \citenamefont {Metz}}]{johnstonVelocityMapImaging2018}%
  \BibitemOpen
  \bibfield  {author} {\bibinfo {author} {\bibfnamefont {M.~D.}\ \bibnamefont {Johnston}}, \bibinfo {author} {\bibfnamefont {W.~L.}\ \bibnamefont {Pearson}, \bibfnamefont {III}}, \bibinfo {author} {\bibfnamefont {G.}~\bibnamefont {Wang}},\ and\ \bibinfo {author} {\bibfnamefont {R.~B.}\ \bibnamefont {Metz}},\ }\href {https://doi.org/10.1063/1.5012896} {\bibfield  {journal} {\bibinfo  {journal} {Review of Scientific Instruments}\ }\textbf {\bibinfo {volume} {89}},\ \bibinfo {pages} {014102} (\bibinfo {year} {2018})}\BibitemShut {NoStop}%
\bibitem [{\citenamefont {Champeau}\ \emph {et~al.}(1994)\citenamefont {Champeau}, \citenamefont {Crubellier}, \citenamefont {Gaardsted}, \citenamefont {Marescaux},\ and\ \citenamefont {Pavolini}}]{champeauKineticEnergyMeasurements1994}%
  \BibitemOpen
  \bibfield  {author} {\bibinfo {author} {\bibfnamefont {R.~J.}\ \bibnamefont {Champeau}}, \bibinfo {author} {\bibfnamefont {A.}~\bibnamefont {Crubellier}}, \bibinfo {author} {\bibfnamefont {J.~O.}\ \bibnamefont {Gaardsted}}, \bibinfo {author} {\bibfnamefont {D.}~\bibnamefont {Marescaux}},\ and\ \bibinfo {author} {\bibfnamefont {D.}~\bibnamefont {Pavolini}},\ }\href {https://doi.org/10.1088/0953-4075/27/5/010} {\bibfield  {journal} {\bibinfo  {journal} {Journal of Physics B: Atomic, Molecular and Optical Physics}\ }\textbf {\bibinfo {volume} {27}},\ \bibinfo {pages} {905} (\bibinfo {year} {1994})}\BibitemShut {NoStop}%
\bibitem [{\citenamefont {Lakhmanskaya}\ \emph {et~al.}(2014)\citenamefont {Lakhmanskaya}, \citenamefont {Best}, \citenamefont {Kumar}, \citenamefont {Endres}, \citenamefont {Hauser}, \citenamefont {Otto}, \citenamefont {Eisenbach}, \citenamefont {{von Zastrow}},\ and\ \citenamefont {Wester}}]{lakhmanskayaPropertiesMultipoleIon2014}%
  \BibitemOpen
  \bibfield  {author} {\bibinfo {author} {\bibfnamefont {O.~Y.}\ \bibnamefont {Lakhmanskaya}}, \bibinfo {author} {\bibfnamefont {T.}~\bibnamefont {Best}}, \bibinfo {author} {\bibfnamefont {S.~S.}\ \bibnamefont {Kumar}}, \bibinfo {author} {\bibfnamefont {E.~S.}\ \bibnamefont {Endres}}, \bibinfo {author} {\bibfnamefont {D.}~\bibnamefont {Hauser}}, \bibinfo {author} {\bibfnamefont {R.}~\bibnamefont {Otto}}, \bibinfo {author} {\bibfnamefont {S.}~\bibnamefont {Eisenbach}}, \bibinfo {author} {\bibfnamefont {A.}~\bibnamefont {{von Zastrow}}},\ and\ \bibinfo {author} {\bibfnamefont {R.}~\bibnamefont {Wester}},\ }\href {https://doi.org/10.1016/j.ijms.2014.03.001} {\bibfield  {journal} {\bibinfo  {journal} {International Journal of Mass Spectrometry}\ }\textbf {\bibinfo {volume} {365--366}},\ \bibinfo {pages} {281} (\bibinfo {year} {2014})}\BibitemShut {NoStop}%
\bibitem [{\citenamefont {Schmid}\ \emph {et~al.}(2022)\citenamefont {Schmid}, \citenamefont {Asvany}, \citenamefont {Salomon}, \citenamefont {Thorwirth},\ and\ \citenamefont {Schlemmer}}]{schmidLeakOutSpectroscopyUniversal2022}%
  \BibitemOpen
  \bibfield  {author} {\bibinfo {author} {\bibfnamefont {P.~C.}\ \bibnamefont {Schmid}}, \bibinfo {author} {\bibfnamefont {O.}~\bibnamefont {Asvany}}, \bibinfo {author} {\bibfnamefont {T.}~\bibnamefont {Salomon}}, \bibinfo {author} {\bibfnamefont {S.}~\bibnamefont {Thorwirth}},\ and\ \bibinfo {author} {\bibfnamefont {S.}~\bibnamefont {Schlemmer}},\ }\href {https://doi.org/10.1021/acs.jpca.2c05767} {\bibfield  {journal} {\bibinfo  {journal} {The Journal of Physical Chemistry A}\ }\textbf {\bibinfo {volume} {126}},\ \bibinfo {pages} {8111} (\bibinfo {year} {2022})}\BibitemShut {NoStop}%
\bibitem [{\citenamefont {Andresen}\ \emph {et~al.}(2010)\citenamefont {Andresen}, \citenamefont {Ashkezari}, \citenamefont {{Baquero-Ruiz}}, \citenamefont {Bertsche}, \citenamefont {Bowe}, \citenamefont {Butler}, \citenamefont {Cesar}, \citenamefont {Chapman}, \citenamefont {Charlton}, \citenamefont {Fajans}, \citenamefont {Friesen}, \citenamefont {Fujiwara}, \citenamefont {Gill}, \citenamefont {Hangst}, \citenamefont {Hardy}, \citenamefont {Hayano}, \citenamefont {Hayden}, \citenamefont {Humphries}, \citenamefont {Hydomako}, \citenamefont {Jonsell}, \citenamefont {Kurchaninov}, \citenamefont {Lambo}, \citenamefont {Madsen}, \citenamefont {Menary}, \citenamefont {Nolan}, \citenamefont {Olchanski}, \citenamefont {Olin}, \citenamefont {Povilus}, \citenamefont {Pusa}, \citenamefont {Robicheaux}, \citenamefont {Sarid}, \citenamefont {Silveira}, \citenamefont {So}, \citenamefont {Storey}, \citenamefont {Thompson}, \citenamefont {{van der Werf}}, \citenamefont {Wilding}, \citenamefont {Wurtele},\ and\ \citenamefont {Yamazaki}}]{Andresen2010}%
  \BibitemOpen
  \bibfield  {author} {\bibinfo {author} {\bibfnamefont {G.~B.}\ \bibnamefont {Andresen}}, \bibinfo {author} {\bibfnamefont {M.~D.}\ \bibnamefont {Ashkezari}}, \bibinfo {author} {\bibfnamefont {M.}~\bibnamefont {{Baquero-Ruiz}}}, \bibinfo {author} {\bibfnamefont {W.}~\bibnamefont {Bertsche}}, \bibinfo {author} {\bibfnamefont {P.~D.}\ \bibnamefont {Bowe}}, \bibinfo {author} {\bibfnamefont {E.}~\bibnamefont {Butler}}, \bibinfo {author} {\bibfnamefont {C.~L.}\ \bibnamefont {Cesar}}, \bibinfo {author} {\bibfnamefont {S.}~\bibnamefont {Chapman}}, \bibinfo {author} {\bibfnamefont {M.}~\bibnamefont {Charlton}}, \bibinfo {author} {\bibfnamefont {J.}~\bibnamefont {Fajans}}, \bibinfo {author} {\bibfnamefont {T.}~\bibnamefont {Friesen}}, \bibinfo {author} {\bibfnamefont {M.~C.}\ \bibnamefont {Fujiwara}}, \bibinfo {author} {\bibfnamefont {D.~R.}\ \bibnamefont {Gill}}, \bibinfo {author} {\bibfnamefont {J.~S.}\ \bibnamefont {Hangst}}, \bibinfo {author} {\bibfnamefont {W.~N.}\ \bibnamefont {Hardy}}, \bibinfo {author} {\bibfnamefont {R.~S.}\ \bibnamefont {Hayano}}, \bibinfo {author} {\bibfnamefont {M.~E.}\ \bibnamefont {Hayden}}, \bibinfo {author} {\bibfnamefont {A.}~\bibnamefont {Humphries}}, \bibinfo {author} {\bibfnamefont {R.}~\bibnamefont {Hydomako}}, \bibinfo {author} {\bibfnamefont {S.}~\bibnamefont {Jonsell}}, \bibinfo {author} {\bibfnamefont {L.}~\bibnamefont {Kurchaninov}}, \bibinfo {author} {\bibfnamefont {R.}~\bibnamefont {Lambo}}, \bibinfo {author} {\bibfnamefont {N.}~\bibnamefont {Madsen}}, \bibinfo {author} {\bibfnamefont {S.}~\bibnamefont {Menary}}, \bibinfo {author} {\bibfnamefont {P.}~\bibnamefont {Nolan}}, \bibinfo {author} {\bibfnamefont {K.}~\bibnamefont {Olchanski}}, \bibinfo {author} {\bibfnamefont {A.}~\bibnamefont {Olin}}, \bibinfo {author} {\bibfnamefont {A.}~\bibnamefont {Povilus}}, \bibinfo {author} {\bibfnamefont {P.}~\bibnamefont {Pusa}}, \bibinfo {author} {\bibfnamefont {F.}~\bibnamefont {Robicheaux}}, \bibinfo {author} {\bibfnamefont
  {E.}~\bibnamefont {Sarid}}, \bibinfo {author} {\bibfnamefont {D.~M.}\ \bibnamefont {Silveira}}, \bibinfo {author} {\bibfnamefont {C.}~\bibnamefont {So}}, \bibinfo {author} {\bibfnamefont {J.~W.}\ \bibnamefont {Storey}}, \bibinfo {author} {\bibfnamefont {R.~I.}\ \bibnamefont {Thompson}}, \bibinfo {author} {\bibfnamefont {D.~P.}\ \bibnamefont {{van der Werf}}}, \bibinfo {author} {\bibfnamefont {D.}~\bibnamefont {Wilding}}, \bibinfo {author} {\bibfnamefont {J.~S.}\ \bibnamefont {Wurtele}},\ and\ \bibinfo {author} {\bibfnamefont {Y.}~\bibnamefont {Yamazaki}} (\bibinfo {collaboration} {ALPHA Collaboration}),\ }\href {https://doi.org/10.1103/PhysRevLett.105.013003} {\bibfield  {journal} {\bibinfo  {journal} {Phys. Rev. Lett.}\ }\textbf {\bibinfo {volume} {105}},\ \bibinfo {pages} {013003} (\bibinfo {year} {2010})}\BibitemShut {NoStop}%
\bibitem [{\citenamefont {Tauch}\ \emph {et~al.}(2023)\citenamefont {Tauch}, \citenamefont {Hassan}, \citenamefont {N{\"o}tzold}, \citenamefont {Endres}, \citenamefont {Wester},\ and\ \citenamefont {Weidem{\"u}ller}}]{Tauch2023}%
  \BibitemOpen
  \bibfield  {author} {\bibinfo {author} {\bibfnamefont {J.}~\bibnamefont {Tauch}}, \bibinfo {author} {\bibfnamefont {S.~Z.}\ \bibnamefont {Hassan}}, \bibinfo {author} {\bibfnamefont {M.}~\bibnamefont {N{\"o}tzold}}, \bibinfo {author} {\bibfnamefont {E.~S.}\ \bibnamefont {Endres}}, \bibinfo {author} {\bibfnamefont {R.}~\bibnamefont {Wester}},\ and\ \bibinfo {author} {\bibfnamefont {M.}~\bibnamefont {Weidem{\"u}ller}},\ }\href {https://doi.org/10.1038/s41567-023-02084-6} {\bibfield  {journal} {\bibinfo  {journal} {Nat. Phys.}\ }\textbf {\bibinfo {volume} {19}},\ \bibinfo {pages} {1270} (\bibinfo {year} {2023})}\BibitemShut {NoStop}%
\bibitem [{\citenamefont {DeVoe}(2009)}]{devoePowerLawDistributionsTrapped2009}%
  \BibitemOpen
  \bibfield  {author} {\bibinfo {author} {\bibfnamefont {R.~G.}\ \bibnamefont {DeVoe}},\ }\href {https://doi.org/10.1103/PhysRevLett.102.063001} {\bibfield  {journal} {\bibinfo  {journal} {Physical Review Letters}\ }\textbf {\bibinfo {volume} {102}},\ \bibinfo {pages} {063001} (\bibinfo {year} {2009})}\BibitemShut {NoStop}%
\bibitem [{\citenamefont {Asvany}\ and\ \citenamefont {Schlemmer}(2009)}]{asvanyNumericalSimulationsKinetic2009}%
  \BibitemOpen
  \bibfield  {author} {\bibinfo {author} {\bibfnamefont {O.}~\bibnamefont {Asvany}}\ and\ \bibinfo {author} {\bibfnamefont {S.}~\bibnamefont {Schlemmer}},\ }\href {https://doi.org/10.1016/j.ijms.2008.10.022} {\bibfield  {journal} {\bibinfo  {journal} {International Journal of Mass Spectrometry}\ }\textbf {\bibinfo {volume} {279}},\ \bibinfo {pages} {147} (\bibinfo {year} {2009})}\BibitemShut {NoStop}%
\bibitem [{\citenamefont {Rajeevan}\ \emph {et~al.}(2021)\citenamefont {Rajeevan}, \citenamefont {Mohandas},\ and\ \citenamefont {Kumar}}]{rajeevanNumericalSimulationsStorage2021}%
  \BibitemOpen
  \bibfield  {author} {\bibinfo {author} {\bibfnamefont {G.}~\bibnamefont {Rajeevan}}, \bibinfo {author} {\bibfnamefont {S.}~\bibnamefont {Mohandas}},\ and\ \bibinfo {author} {\bibfnamefont {S.~S.}\ \bibnamefont {Kumar}},\ }\href {https://doi.org/10.1088/1402-4896/ac1472} {\bibfield  {journal} {\bibinfo  {journal} {Physica Scripta}\ }\textbf {\bibinfo {volume} {96}},\ \bibinfo {pages} {124001} (\bibinfo {year} {2021})}\BibitemShut {NoStop}%
\bibitem [{\citenamefont {H{\"o}ltkemeier}\ \emph {et~al.}(2016{\natexlab{a}})\citenamefont {H{\"o}ltkemeier}, \citenamefont {Weckesser}, \citenamefont {{L{\'o}pez-Carrera}},\ and\ \citenamefont {Weidem{\"u}ller}}]{holtkemeierBufferGasCoolingSingle2016}%
  \BibitemOpen
  \bibfield  {author} {\bibinfo {author} {\bibfnamefont {B.}~\bibnamefont {H{\"o}ltkemeier}}, \bibinfo {author} {\bibfnamefont {P.}~\bibnamefont {Weckesser}}, \bibinfo {author} {\bibfnamefont {H.}~\bibnamefont {{L{\'o}pez-Carrera}}},\ and\ \bibinfo {author} {\bibfnamefont {M.}~\bibnamefont {Weidem{\"u}ller}},\ }\href {https://doi.org/10.1103/PhysRevLett.116.233003} {\bibfield  {journal} {\bibinfo  {journal} {Physical Review Letters}\ }\textbf {\bibinfo {volume} {116}},\ \bibinfo {pages} {233003} (\bibinfo {year} {2016}{\natexlab{a}})}\BibitemShut {NoStop}%
\bibitem [{\citenamefont {H{\"o}ltkemeier}\ \emph {et~al.}(2016{\natexlab{b}})\citenamefont {H{\"o}ltkemeier}, \citenamefont {Weckesser}, \citenamefont {{L{\'o}pez-Carrera}},\ and\ \citenamefont {Weidem{\"u}ller}}]{holtkemeierDynamicsSingleTrapped2016}%
  \BibitemOpen
  \bibfield  {author} {\bibinfo {author} {\bibfnamefont {B.}~\bibnamefont {H{\"o}ltkemeier}}, \bibinfo {author} {\bibfnamefont {P.}~\bibnamefont {Weckesser}}, \bibinfo {author} {\bibfnamefont {H.}~\bibnamefont {{L{\'o}pez-Carrera}}},\ and\ \bibinfo {author} {\bibfnamefont {M.}~\bibnamefont {Weidem{\"u}ller}},\ }\href {https://doi.org/10.1103/PhysRevA.94.062703} {\bibfield  {journal} {\bibinfo  {journal} {Physical Review A}\ }\textbf {\bibinfo {volume} {94}},\ \bibinfo {pages} {062703} (\bibinfo {year} {2016}{\natexlab{b}})}\BibitemShut {NoStop}%
\bibitem [{\citenamefont {Svendsen}\ \emph {et~al.}(2013)\citenamefont {Svendsen}, \citenamefont {Lammich}, \citenamefont {Andersen}, \citenamefont {Bechtold}, \citenamefont {S{\o}ndergaard}, \citenamefont {Mikkelsen},\ and\ \citenamefont {Pedersen}}]{svendsenTrappingIonsFast2013}%
  \BibitemOpen
  \bibfield  {author} {\bibinfo {author} {\bibfnamefont {A.}~\bibnamefont {Svendsen}}, \bibinfo {author} {\bibfnamefont {L.}~\bibnamefont {Lammich}}, \bibinfo {author} {\bibfnamefont {J.~E.}\ \bibnamefont {Andersen}}, \bibinfo {author} {\bibfnamefont {H.~K.}\ \bibnamefont {Bechtold}}, \bibinfo {author} {\bibfnamefont {E.}~\bibnamefont {S{\o}ndergaard}}, \bibinfo {author} {\bibfnamefont {F.}~\bibnamefont {Mikkelsen}},\ and\ \bibinfo {author} {\bibfnamefont {H.~B.}\ \bibnamefont {Pedersen}},\ }\href {https://doi.org/10.1103/PhysRevA.87.043410} {\bibfield  {journal} {\bibinfo  {journal} {Physical Review A}\ }\textbf {\bibinfo {volume} {87}},\ \bibinfo {pages} {043410} (\bibinfo {year} {2013})}\BibitemShut {NoStop}%
\bibitem [{\citenamefont {F{\"u}rst}\ \emph {et~al.}(2018)\citenamefont {F{\"u}rst}, \citenamefont {Ewald}, \citenamefont {Secker}, \citenamefont {Joger}, \citenamefont {Feldker},\ and\ \citenamefont {Gerritsma}}]{furstProspectsReachingQuantum2018}%
  \BibitemOpen
  \bibfield  {author} {\bibinfo {author} {\bibfnamefont {H.~A.}\ \bibnamefont {F{\"u}rst}}, \bibinfo {author} {\bibfnamefont {N.~V.}\ \bibnamefont {Ewald}}, \bibinfo {author} {\bibfnamefont {T.}~\bibnamefont {Secker}}, \bibinfo {author} {\bibfnamefont {J.}~\bibnamefont {Joger}}, \bibinfo {author} {\bibfnamefont {T.}~\bibnamefont {Feldker}},\ and\ \bibinfo {author} {\bibfnamefont {R.}~\bibnamefont {Gerritsma}},\ }\href {https://doi.org/10.1088/1361-6455/aadd7d} {\bibfield  {journal} {\bibinfo  {journal} {Journal of Physics B: Atomic, Molecular and Optical Physics}\ }\textbf {\bibinfo {volume} {51}},\ \bibinfo {pages} {195001} (\bibinfo {year} {2018})}\BibitemShut {NoStop}%
\bibitem [{\citenamefont {Trimby}\ \emph {et~al.}(2022)\citenamefont {Trimby}, \citenamefont {Hirzler}, \citenamefont {F{\"u}rst}, \citenamefont {{Safavi-Naini}}, \citenamefont {Gerritsma},\ and\ \citenamefont {Lous}}]{trimbyBufferGasCooling2022}%
  \BibitemOpen
  \bibfield  {author} {\bibinfo {author} {\bibfnamefont {E.}~\bibnamefont {Trimby}}, \bibinfo {author} {\bibfnamefont {H.}~\bibnamefont {Hirzler}}, \bibinfo {author} {\bibfnamefont {H.}~\bibnamefont {F{\"u}rst}}, \bibinfo {author} {\bibfnamefont {A.}~\bibnamefont {{Safavi-Naini}}}, \bibinfo {author} {\bibfnamefont {R.}~\bibnamefont {Gerritsma}},\ and\ \bibinfo {author} {\bibfnamefont {R.~S.}\ \bibnamefont {Lous}},\ }\href {https://doi.org/10.1088/1367-2630/ac5759} {\bibfield  {journal} {\bibinfo  {journal} {New Journal of Physics}\ }\textbf {\bibinfo {volume} {24}},\ \bibinfo {pages} {035004} (\bibinfo {year} {2022})}\BibitemShut {NoStop}%
\bibitem [{\citenamefont {Jusko}\ \emph {et~al.}(2024)\citenamefont {Jusko}, \citenamefont {{Jim{\'e}nez-Redondo}},\ and\ \citenamefont {Caselli}}]{juskoColdCASIon2023}%
  \BibitemOpen
  \bibfield  {author} {\bibinfo {author} {\bibfnamefont {P.}~\bibnamefont {Jusko}}, \bibinfo {author} {\bibfnamefont {M.}~\bibnamefont {{Jim{\'e}nez-Redondo}}},\ and\ \bibinfo {author} {\bibfnamefont {P.}~\bibnamefont {Caselli}},\ }\href {https://doi.org/10.1080/00268976.2023.2217744} {\bibfield  {journal} {\bibinfo  {journal} {Molecular Physics}\ }\textbf {\bibinfo {volume} {122}},\ \bibinfo {pages} {e2217744} (\bibinfo {year} {2024})}\BibitemShut {NoStop}%
\bibitem [{\citenamefont {Gerlich}(1992)}]{gerlichInhomogeneousRFFields1992}%
  \BibitemOpen
  \bibfield  {author} {\bibinfo {author} {\bibfnamefont {D.}~\bibnamefont {Gerlich}},\ }in\ \href {https://doi.org/10.1002/9780470141397.ch1} {\emph {\bibinfo {booktitle} {Advances in {{Chemical Physics}}}}}\ (\bibinfo  {publisher} {{John Wiley \& Sons, Ltd}},\ \bibinfo {year} {1992})\ pp.\ \bibinfo {pages} {1--176}\BibitemShut {NoStop}%
\bibitem [{\citenamefont {Swope}\ \emph {et~al.}(1982)\citenamefont {Swope}, \citenamefont {Andersen}, \citenamefont {Berens},\ and\ \citenamefont {Wilson}}]{swopeComputerSimulationMethod1982}%
  \BibitemOpen
  \bibfield  {author} {\bibinfo {author} {\bibfnamefont {W.~C.}\ \bibnamefont {Swope}}, \bibinfo {author} {\bibfnamefont {H.~C.}\ \bibnamefont {Andersen}}, \bibinfo {author} {\bibfnamefont {P.~H.}\ \bibnamefont {Berens}},\ and\ \bibinfo {author} {\bibfnamefont {K.~R.}\ \bibnamefont {Wilson}},\ }\href {https://doi.org/10.1063/1.442716} {\bibfield  {journal} {\bibinfo  {journal} {The Journal of Chemical Physics}\ }\textbf {\bibinfo {volume} {76}},\ \bibinfo {pages} {637} (\bibinfo {year} {1982})}\BibitemShut {NoStop}%
\bibitem [{\citenamefont {Ruth}(1983)}]{ruthCanonicalIntegrationTechnique1983}%
  \BibitemOpen
  \bibfield  {author} {\bibinfo {author} {\bibfnamefont {R.~D.}\ \bibnamefont {Ruth}},\ }\href {https://doi.org/10.1109/TNS.1983.4332919} {\bibfield  {journal} {\bibinfo  {journal} {IEEE Transactions on Nuclear Science}\ }\textbf {\bibinfo {volume} {30}},\ \bibinfo {pages} {2669} (\bibinfo {year} {1983})}\BibitemShut {NoStop}%
\bibitem [{\citenamefont {{\'S}migaj}\ \emph {et~al.}(2015)\citenamefont {{\'S}migaj}, \citenamefont {Betcke}, \citenamefont {Arridge}, \citenamefont {Phillips},\ and\ \citenamefont {Schweiger}}]{smigajSolvingBoundaryIntegral2015}%
  \BibitemOpen
  \bibfield  {author} {\bibinfo {author} {\bibfnamefont {W.}~\bibnamefont {{\'S}migaj}}, \bibinfo {author} {\bibfnamefont {T.}~\bibnamefont {Betcke}}, \bibinfo {author} {\bibfnamefont {S.}~\bibnamefont {Arridge}}, \bibinfo {author} {\bibfnamefont {J.}~\bibnamefont {Phillips}},\ and\ \bibinfo {author} {\bibfnamefont {M.}~\bibnamefont {Schweiger}},\ }\href {https://doi.org/10.1145/2590830} {\bibfield  {journal} {\bibinfo  {journal} {ACM Transactions on Mathematical Software}\ }\textbf {\bibinfo {volume} {41}},\ \bibinfo {pages} {6:1} (\bibinfo {year} {2015})}\BibitemShut {NoStop}%
\bibitem [{\citenamefont {Betcke}\ and\ \citenamefont {Scroggs}(2021)}]{betckeBemppclFastPython2021}%
  \BibitemOpen
  \bibfield  {author} {\bibinfo {author} {\bibfnamefont {T.}~\bibnamefont {Betcke}}\ and\ \bibinfo {author} {\bibfnamefont {M.~W.}\ \bibnamefont {Scroggs}},\ }\href {https://doi.org/10.21105/joss.02879} {\bibfield  {journal} {\bibinfo  {journal} {Journal of Open Source Software}\ }\textbf {\bibinfo {volume} {6}},\ \bibinfo {pages} {2879} (\bibinfo {year} {2021})}\BibitemShut {NoStop}%
\bibitem [{Sal()}]{SalomePlatform}%
  \BibitemOpen
  \href@noop {} {\bibinfo {title} {Salome platform}},\ \bibinfo {howpublished} {https://www.salome-platform.org/}\BibitemShut {NoStop}%
\bibitem [{\citenamefont {Londry}\ \emph {et~al.}(1993)\citenamefont {Londry}, \citenamefont {Alfred},\ and\ \citenamefont {March}}]{londryComputerSimulationSingleion1993}%
  \BibitemOpen
  \bibfield  {author} {\bibinfo {author} {\bibfnamefont {F.~A.}\ \bibnamefont {Londry}}, \bibinfo {author} {\bibfnamefont {R.~L.}\ \bibnamefont {Alfred}},\ and\ \bibinfo {author} {\bibfnamefont {R.~E.}\ \bibnamefont {March}},\ }\href {https://doi.org/10.1016/1044-0305(93)80047-3} {\bibfield  {journal} {\bibinfo  {journal} {Journal of the American Society for Mass Spectrometry}\ }\textbf {\bibinfo {volume} {4}},\ \bibinfo {pages} {687} (\bibinfo {year} {1993})}\BibitemShut {NoStop}%
\bibitem [{\citenamefont {Tsallis}(1988)}]{tsallisPossibleGeneralizationBoltzmannGibbs1988}%
  \BibitemOpen
  \bibfield  {author} {\bibinfo {author} {\bibfnamefont {C.}~\bibnamefont {Tsallis}},\ }\href {https://doi.org/10.1007/BF01016429} {\bibfield  {journal} {\bibinfo  {journal} {Journal of Statistical Physics}\ }\textbf {\bibinfo {volume} {52}},\ \bibinfo {pages} {479} (\bibinfo {year} {1988})}\BibitemShut {NoStop}%
\bibitem [{\citenamefont {Gerlich}(1990)}]{gerlichOrthoParaTransitions1990}%
  \BibitemOpen
  \bibfield  {author} {\bibinfo {author} {\bibfnamefont {D.}~\bibnamefont {Gerlich}},\ }\href {https://doi.org/10.1063/1.457980} {\bibfield  {journal} {\bibinfo  {journal} {The Journal of Chemical Physics}\ }\textbf {\bibinfo {volume} {92}},\ \bibinfo {pages} {2377} (\bibinfo {year} {1990})}\BibitemShut {NoStop}%
\bibitem [{\citenamefont {Bast}\ \emph {et~al.}(2023)\citenamefont {Bast}, \citenamefont {B{\"o}ing}, \citenamefont {Salomon}, \citenamefont {Thorwirth}, \citenamefont {Asvany}, \citenamefont {Sch{\"a}fer},\ and\ \citenamefont {Schlemmer}}]{Bast2023}%
  \BibitemOpen
  \bibfield  {author} {\bibinfo {author} {\bibfnamefont {M.}~\bibnamefont {Bast}}, \bibinfo {author} {\bibfnamefont {J.}~\bibnamefont {B{\"o}ing}}, \bibinfo {author} {\bibfnamefont {T.}~\bibnamefont {Salomon}}, \bibinfo {author} {\bibfnamefont {S.}~\bibnamefont {Thorwirth}}, \bibinfo {author} {\bibfnamefont {O.}~\bibnamefont {Asvany}}, \bibinfo {author} {\bibfnamefont {M.}~\bibnamefont {Sch{\"a}fer}},\ and\ \bibinfo {author} {\bibfnamefont {S.}~\bibnamefont {Schlemmer}},\ }\href {https://doi.org/10.1016/j.jms.2023.111840} {\bibfield  {journal} {\bibinfo  {journal} {Journal of Molecular Spectroscopy}\ }\textbf {\bibinfo {volume} {398}},\ \bibinfo {pages} {111840} (\bibinfo {year} {2023})}\BibitemShut {NoStop}%
\end{thebibliography}
\end{document}